\documentclass[aps,prl,twocolumn,superscriptaddress,longbibliography]{revtex4-2}

\usepackage{amssymb,amsmath}
\usepackage{graphicx}
\usepackage{xcolor}
\usepackage{hyperref}
\usepackage{booktabs}

\hypersetup{
    unicode=false,
    pdftoolbar=true,
    pdfmenubar=true,
    pdffitwindow=false,
    pdfstartview={FitH},
    pdfauthor={Jack Diab},
    colorlinks=true,
    linkcolor=blue,
    citecolor=red,
}

\newcommand{\addAA}[1]{\textcolor{black}{#1}}

\begin{document}

\title{Unidirectional Dark-to-Bright Rescue in Cavity-Coupled Quantum Transport}

\author{Jack Diab}
\affiliation{Department of Chemistry and Biochemistry, University of California, Los Angeles, CA 90095, USA}
\author{Arpit Arora}
\email{ararora@nvidia.com}
\affiliation{Division of Physical Sciences, College of Letters and Science, University of California, Los Angeles (UCLA), Los Angeles, CA, USA}
\affiliation{Department of Electrical and Computer Engineering, UCLA, Los Angeles, CA, USA}
\affiliation{NVIDIA Corporation, Santa Clara, CA, USA}
\author{Taylor L. Patti}
\affiliation{NVIDIA Corporation, Santa Clara, CA, USA}
\author{Prineha Narang}
\email{prineha@ucla.edu}
\affiliation{Division of Physical Sciences, College of Letters and Science, University of California, Los Angeles (UCLA), Los Angeles, CA, USA}
\affiliation{Department of Electrical and Computer Engineering, UCLA, Los Angeles, CA, USA}

\begin{abstract}
Strong light–matter coupling in optical microcavities can transport energy ballistically across an emitter array, but the same coupling buries most of the excitation in a manifold of dark states that grows with system size and traps energy outside the transport channel. We show that the off-diagonal (non-Condon) part of the exciton–phonon coupling opens a one-way escape route from this trap, driving population irreversibly from dark states into the radiative channel. This rate is fixed by a photonic-weight conservation law rather than by dark–bright overlap which evacuates the dark manifold at a rate independent of system size. The mechanism contributes to transport with near-complete efficiency with four signatures being single-exponential dark-state decay, a size-scaling efficiency gap, distinct temperature behavior, and a resonance in the escape rate at vibrational bath modes. Beyond polariton transport, it recasts dark states from a parasitic loss channel into an engineered dissipative resource, with implications for light harvesting and dissipation-based quantum control.
\end{abstract}

\maketitle

Strong \addAA{light-matter} coupling between emitters and a confined optical mode \addAA{can create delocalized polaritonic states} that can transport energy ballistically across an array~\cite{Flick18,RibeiroChemSci2018,Schachenmayer2015,FeistGarciaVidal2015,Orgiu2015,Zhong2016,low2017polaritons,rubies2022photon,herrera2025moire,bittorf2026long}.
\addAA{The same collective coupling, however, generates $N{-}1$ optically dark eigenstates per cavity mode which hold most of the injected energy but cannot release it into the cavity~\cite{Houdre1996,Agranovich2003,Kena-Cohen2019,DelPo2020,Pandya2022,Khazanov2023}}.
Mitigating dark-state trapping is an open and central challenge in polariton transport \addAA{and cavity-mediated energy transfer}~\cite{Sandik2024,Mandal2023,Ribeiro2022,DuChemSci2018,Blazquez18}.

\begin{figure}[!t]
\centering
\includegraphics[width=\columnwidth]{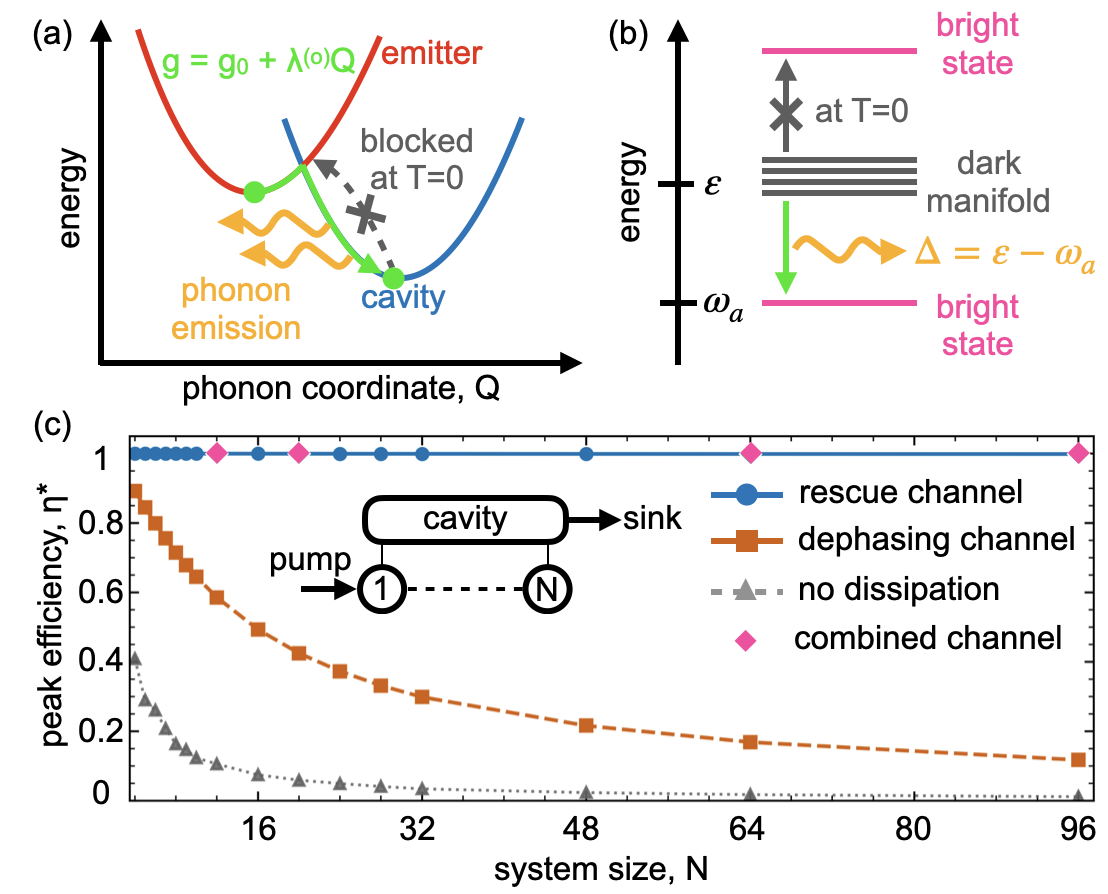}
\caption{Unidirectional dark-to-bright rescue channel.
(a) The non-Condon exchange between quantum emitter and phononic bath modifies the light-matter coupling making phonon emission and absorption causal with transitions in quantum system which opens a one-way energy dump for the quantum system via phonon emission. (b) This implies a polaritonic pathway for relxation from dark manifold into bright state which is unidirectional due to vanishing cavity weight of dark states.
(c) The rescue channel leads to transport independent of system size shown here via the peak efficiency $\eta^*$ vs.\ system size $N$ (cavity drain, $\gamma_{\mathrm{rec}}$ and $\gamma_{\mathrm{deph}}$ optimized per $N$);
$g=1.5$ meV, avergae over 15 disorder realizations.}
\label{fig:schematic}
\end{figure}

In this work, we show a one-way rescue channel from the dark states trap, mediated by bath. This is in contrast to existing approaches to this problem that share a common structure where they enhance transport by promoting equilibration between dark and bright sectors \cite{Sandik2024, Mandal2023, Ribeiro2022}. Pure dephasing from the diagonal (Condon) exciton-phonon coupling provides the standard environment-assisted quantum transport (ENAQT) pathway~\cite{Plenio2008,Rebentrost2009,Cao2009,Lambert2013,Cao2020,Xiang2019,PerezSanchez2025}. 
Coherent re-coupling of dark and bright manifolds has been pursued via selective polariton pumping~\cite{Michail2024} and Floquet driving~\cite{Hubener2021}.
Disorder engineering can endow dark states with partial photonic character~\cite{Gonzalez-Ballestero2016}. In all such cases, transport remains limited by bidirectional dark-bright polaritonic exchange, whose efficiency degrades as the dark manifold grows. A fundamentally different possibility is an irreversible escape channel out of the dark sector itself.


We identify 
unidirectional dark-to-bright rescue driven by the off-diagonal (non-Condon) contribution to coupling between quantum and bath, see Fig.~\ref{fig:schematic}a.
We consider an array of $N$ two level systems where the rescue channel 
transfers excitation from emitter states into the cavity without the reverse process at zero temperature. The channel generates finite transition rates out of the dark manifold while transitions into dark states vanish identically because dark states carry zero photonic weight, see Fig.~\ref{fig:schematic}b. The resulting dynamics therefore acts as a one-way valve from dark to bright sectors, structurally distinct from dephasing-based equilibration; latter is bidirectional in dark and bright state transitions.
While dephasing-mediated transport slows as the dark fraction grows, the rescue process proposed here is protected by photonic-weight sum rule and evacuates the dark manifold at a rate independent of system size, see Fig.~\ref{fig:schematic}c.
The known dissipative processes have appeared in other settings: second-order rate theories built on the Condon coupling yield bidirectional dark–bright transfer suppressed as $1/N$~\cite{Xiang2019,PerezSanchez2025}; polaron-dressed, phonon-mediated cavity feeding is well established for single quantum dots~\cite{Majumdar2011,RoyHughes2011}, where no collective dark manifold exists; and atomistic polariton simulations contain non-Condon effects only implicitly~\cite{Sokolovskii2023,Groenhof2019,Davidsson2023}. None of these provides a one-way escape channel whose rate survives the growth of the dark manifold.
Our framework further predicts experimentally distinguishable signatures including exponential dark-state decay, characteristic size scaling, temperature dependence governed by detailed balance, and resonant enhancement at vibrational bath modes.

\paragraph*{Unidirectional rescue channel:}
We consider $N$ two-level emitters, each with ground state $|g\rangle_i$ and excited state $|e\rangle_i$ and transition energy $\varepsilon$. The emitters are coupled to a single-mode cavity with energy $\omega_a$. The system is described by the Tavis-Cummings-Hubbard Hamiltonian~\cite{ashida2021cavity}
\begin{equation}
\label{eq:H}
\begin{split}
\hat{H} = \omega_a\hat{a}^\dagger\hat{a} + \varepsilon\sum_{i=1}^{N}\hat{\sigma}_i^\dagger\hat{\sigma}_i &+ \sum_{i=1}^{N} g\!\left(\hat{a}^\dagger\hat{\sigma}_i + \mathrm{h.c.}\right) \\
&+ \sum_{i=1}^{N-1} t_i\!\left(\hat{\sigma}_i^\dagger\hat{\sigma}_{i+1} + \mathrm{h.c.}\right),
\end{split}
\end{equation}
where $\hat{\sigma}_i \equiv \hat{\sigma}_i^- = |g\rangle_i\langle e|$ is the exciton lowering operator at site $i$, $\hat{a}$ is the cavity annihilation operator, the cavity coupling $g$ is uniform, and $\Delta = \varepsilon - \omega_a$ is the site-cavity detuning that sets the emitted-phonon energy. We introduce static hopping disorder such that $t_i = t + \delta t\,X_i$~\cite{Anderson1958,Houdre1996,Agranovich2003}, $X_i \sim \mathcal{N}(0,1)$, see SM~\cite{SuppMaterial}).
Diagonalizing $\hat{H}$ gives eigenstate $|\psi_k\rangle$ with cavity weight $w_k  = |\langle \text{cav}|\psi_k\rangle|^2$: two bright polaritons with $w_k\neq 0$ and $N-1$ dark modes which under uniform coupling satisfy $w_k = 0$. Disorder lifts the latter to small but finite $w_k\ll 1$. We work in the single-excitation manifold $\{|\text{cav}\rangle, |1\rangle, \ldots, |N\rangle\}$ as Eq.~(\ref{eq:H}) conserves total excitation number. Here $|{\rm cav}\rangle$ denotes the state with one cavity photon and all emitters in their ground states, and $|i\rangle$ ($i = 1,\ldots,N$) the state with emitter $i$ excited and the cavity empty. 




Each emitter further couples to a phonon bath $\sum_k \omega_k\hat{b}_{i,k}^\dagger  \hat{b}_{i,k}$. To linear order system-bath coupling yields two distinct effects: $\varepsilon\rightarrow \varepsilon + \sum_k\lambda_{i,k}^{(d)}\hat{Q}_{i,k}$ due to diagonal (Condon) coupling \cite{Xiang2019,PerezSanchez2025}, and alteration of light-matter coupling $g \rightarrow g + \sum_k\lambda_{k}^{(o)}\hat{Q}_{i,k}$ due to off-diagonal (non-Condon) coupling $\lambda^{(o)}_k$ \cite{Seibt2018,Kano2002,Takahashi2022,Odewale2024}. These couplings can be computed using first principle calculations~\cite{flick2018cavity,flick2019excited,wang2021light}. Importantly, phonon emission and absorption during non-Condon exchange act as causal agents for transitions in the quantum system which as we show below is the key to dark-to-bright rescue, also see Fig.~\ref{fig:schematic}a.  

Within interaction picture and rotating-wave approximation, we calculate the rates for phonon emission ($\hat{a}^\dagger \hat{\sigma}_i \hat{b}_k^\dagger$) and absorption ($\hat{a} \hat{\sigma}_i^\dagger \hat{b}_k$) processes
\begin{equation}
\gamma^{\mathrm{em}}_{\Delta,T} = 2J(\Delta)[1+n_{\Delta,T}],\quad
\gamma^{\mathrm{abs}}_{\Delta,T} = 2J(\Delta)\,n_{\Delta,T}
\label{eq:rates}
\end{equation}
respectively, with $J(\omega) = (\pi/2)\sum_k |\lambda_k^{(o)}|^2\delta(\omega - \omega_k)$ being the non-Condon spectral density, and $\Delta = \varepsilon - \omega_a>0$; $n_{\omega,T}$ is the Bose occupation function (see SM for details). Note that as $T\rightarrow 0$, $\gamma^{\rm abs}\rightarrow 0$ and $\gamma^{\rm em}\rightarrow 2J(\Delta)$, which implies that non-Condon exchange opens a one-way energy dump for light-matter hybrid to lower its energy from dark state manifold to lower bright state, see Fig.~\ref{fig:schematic}b.

We next consider the effect of $\gamma_{\rm em, abs}$ on the dynamics of the open-quantum system, $\dot{\hat{\rho}} = -i[\hat{H},\hat{\rho}] + \mathcal{D}[\hat{\rho}]$. Here, the dissipator $\mathcal{D}[\hat{\rho}] = \sum_\mu \big(\hat{L}_\mu\hat{\rho}\hat{L}_\mu^\dagger - \tfrac{1}{2}\{\hat{L}_\mu^\dagger\hat{L}_\mu,\hat{\rho}\}\big)$, where $\mu$ runs over three families of jump operators. We have (i) pure dephasing, $\hat{L}_{{\rm deph},i} = \sqrt{\gamma_{\rm deph}}\,\hat{\sigma}_i^\dagger\hat{\sigma}_i$, generated by the Condon exchange and central to the ENAQT formulation \cite{Plenio2008,Rebentrost2009,Cao2009}; (ii) the non-Condon emission and absorption channels of Eq.~(\ref{eq:rates}); and (iii) a drain $\hat{L}_{\rm drain}$ that extracts the excitation, specified in the transport configuration below. At $T\rightarrow 0$ the absorption process switches off, and channel (ii) reduces to the single jump operator
\begin{equation}
\label{eq:Lrec}
\hat{L}_{{\rm rec},i} = \sqrt{\gamma_{\rm rec}}\,\hat{a}^\dagger\hat{\sigma}_i,
\qquad
\gamma_{\rm rec} \equiv \gamma^{\rm em}_{\Delta,T=0} = 2J(\Delta).
\end{equation}
Projecting onto the eigenbasis $\{|\psi_k\rangle\}$, $\hat{L}_{{\rm rec},i}$ drives transitions $|\psi_l\rangle \rightarrow |\psi_k\rangle$ with secular rate $W^{({\rm rec})}_{k\leftarrow l} = \gamma_{\rm rec} \sum_i |\langle \psi_k | \hat{a}^\dagger \hat{\sigma}_i|\psi_l\rangle|^2$, which in the single-excitation manifold reduces to
\begin{equation}
\label{eq:rescue}
    W^{({\rm rec})}_{k\leftarrow l} = \gamma_{\rm rec}\, w_k(1 - w_l).
\end{equation}
Importantly, $w_k = 0$ throughout the dark manifold. Thus, transitions into dark states vanish identically\addAA{, establishing Eq.~(\ref{eq:rescue}) as a strictly unidirectional rate. Furthermore,} the photonic-weight sum rule $\sum_{k\in\mathcal{B}}w_k = 1$ then fixes the total rate out of any dark state to $\sum_{k\in\mathcal{B}} W_{k\leftarrow l\in\mathcal{D}}^{(\mathrm{rec})} = \gamma_{\mathrm{rec}}$, independent of the dark state's energy, spatial structure, and crucially system size, $N$.
The rescue channel is therefore a strict one-way valve \addAA{which neither dilutes with system size nor depends on which dark state is occupied}. \addAA{As we show in SM,} unidirectionality is a property of the full eigenbasis Redfield tensor, not of the secular approximation, and holds for any bath spectral density at $T=0$~\cite{SuppMaterial}. \addAA{The dephasing channel (i), by contrast, generates the rate} $W_{k\leftarrow l}^{(\mathrm{deph})} = \gamma_{\rm deph}\sum_i|\langle \psi_k | \sigma_i^\dagger \sigma_i|\psi_l\rangle|^2$, \addAA{which} is symmetric in $k,l$ and therefore bidirectional.

\begin{figure*}[t]
\centering
\includegraphics[width=\textwidth]{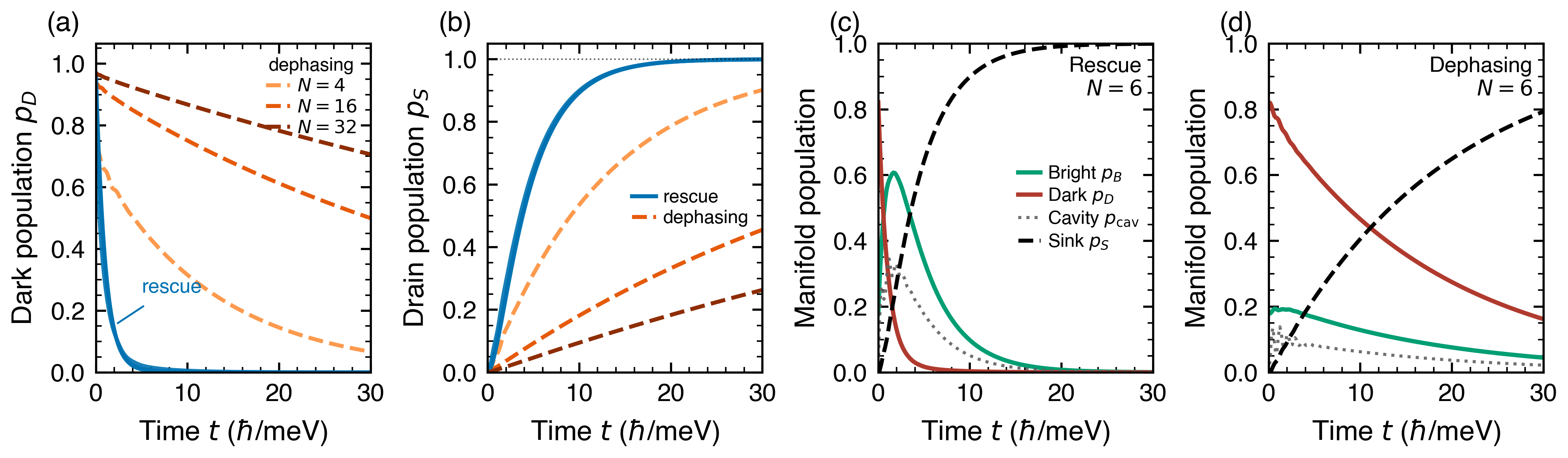}
\caption{Cavity-drain transport dynamics at matched rates $\gamma_{\mathrm{rec}} = \gamma_{\mathrm{deph}} = 1.0$ meV, $g = 1.5$ meV, $\delta t = 0.5$ meV, $\gamma_{\mathrm{lead}} = 0.5$ meV; ensemble-averaged over \addAA{15 disorder realizations in (a),(b) and 20 in (c),(d)}.
(a) Dark-manifold population and (b) cumulative drain population (transport efficiency $\eta = p_S$) vs.\ time for system sizes $N=4,16,32$: \addAA{the rescue curves (blue, solid) collapse onto a single exponential with $\eta\to1$ at every $N$, while dephasing (orange, dashed; darker with increasing $N$)} leaves a long-lived dark plateau and a size-dependent efficiency deficit.
(c),(d) Manifold-resolved populations at $N=6$ under (c) pure rescue and (d) pure dephasing: bright $p_B$ (green), dark $p_D$ (red), cavity $p_{\mathrm{cav}}$ (gray dotted), and sink $p_S$ (black dashed). The transient bright peak in (c) is the distinctive signature of the unidirectional valve\addAA{: population passes \emph{through} the bright manifold on its way out of the dark reservoir; under dephasing the bright manifold drains first and $\eta$ saturates below unity}.}
\label{fig:dynamics}
\end{figure*}

\paragraph*{Cavity-drain regime.}
The structural unidirectionality of the rescue channel translates into a quantitative finite-time transport advantage in the cavity-drain configuration. \addAA{ In the configuration relevant to polariton experiments~\cite{Orgiu2015,Zhong2016,Hutchison2012} the excitation is extracted as a cavity photon. The drain channel (iii) takes the form $\hat{L}_{\rm drain} = \sqrt{\gamma_{\rm lead}}|S\rangle\langle {\rm cav}|$, where $|S\rangle$ is a sink state that accumulates the extracted population. This channel empties $|\psi_k\rangle$ at rate $\gamma_{\rm lead}w_k$ which along with $w_{k\in D} = 0$ in a clean system leads to saturated transport efficiency, $\eta \sim 1/N$. The rescue channel governed by $W_{\rm rec}\neq 0$ fixed by photonic sum rule provides an alternative approach of enhancing cavity transport which is independent of system size, see Eq.~(\ref{eq:rescue}).}


\addAA{The excitation is loaded at the source, and evolved under the full Lindblad equation. Rescue (dark $\rightarrow$ cavity, rate $\gamma_{\rm rec}$) followed by extraction (cavity $\rightarrow$ sink, rate $\gamma_{\rm lead}$) forms a one-way cascade, against which pure dephasing supplies only a bidirectional exchange (dark $\leftrightarrow$ bright, rate $\gamma_{\rm deph}$)
.} The lumped rate equations in this configuration obey
\begin{align}
\dot{p}_B &= -\gamma_{\mathrm{lead}} p_B + \gamma_{\mathrm{rec}}\,p_D + \gamma_{\mathrm{deph}}\Lambda(p_D - p_B), \nonumber \\
\dot{p}_D &= -\gamma_{\mathrm{rec}}\,p_D - \gamma_{\mathrm{deph}}\Lambda(p_D - p_B), \nonumber \\
\dot{p}_S &= \gamma_{\mathrm{lead}}\,p_B,
\label{eq:rate_eqs}
\end{align}
where $p_{B(D)}$ is the bright (dark) population, $p_S$ the drained sink population, and $\Lambda \sim 1/N$ a geometric constant. The cavity is the only radiative port, the sole state carrying photonic weight, so its decay into the sink at rate $\gamma_{\rm lead}$ is the experimentally observable cavity emission, and the cumulative sink population defines the transport efficiency, $\eta \equiv p_S$. For pure rescue ($\gamma_{\mathrm{deph}}=0$) the dark population decays exponentially, $p_D(t) = p_D(0)\,e^{-\gamma_{\mathrm{rec}} t}$, and the rate equations~(\ref{eq:rate_eqs}) integrate to a closed-form recycling efficiency $\eta(t) = 1 - A\,e^{-\gamma_{\mathrm{lead}} t} - B\,e^{-\gamma_{\mathrm{rec}} t}$ (where $A+B=1$ is derived in the SM~\cite{SuppMaterial}); hence \addAA{$\eta \rightarrow 1$} on the timescale $\max(\gamma_{\mathrm{rec}}^{-1}, \gamma_{\mathrm{lead}}^{-1})$, with no residual dependence on $N$ because $\gamma_{\mathrm{rec}}$ is fixed by the photonic-weight sum rule.
For pure dephasing ($\gamma_{\mathrm{rec}}=0$), the dark population first relaxes to the symmetric equilibrium $p_D \to p_B$ (limited by $\gamma_{\mathrm{deph}}\Lambda$), then drains only by leaking through the bright channel.
At any finite measurement time $t_{\rm mes}$, dephasing therefore delivers a smaller drained fraction than rescue, with the gap controlled by $\gamma_{\mathrm{deph}}\Lambda t_{\rm mes}$.

We test this prediction numerically by direct propagation of the full Lindblad equation in the extended Hilbert space (cavity + $N$ sites + sink).
Fig.~\ref{fig:dynamics}(a,b) confirms the predicted dynamics across system sizes $N=4,16,32$. Under rescue the dark population decays exponentially and the efficiency reaches unity independently of $N$, whereas under dephasing the residual dark population and the efficiency deficit both grow with $N$.

\paragraph*{Manifold-resolved valve dynamics.}
Resolving the populations into the bright, dark, cavity, and sink manifolds visualizes the valve mechanism directly.
Fig.~\ref{fig:dynamics}(c,d) track all four projectors.
Under pure rescue [Fig.~\ref{fig:dynamics}(c)], the dark manifold evacuates exponentially while a \emph{transient} bright population peaking near $t \sim \gamma_{\mathrm{rec}}^{-1}$ funnels the excitation into the cavity drain; the sink rises smoothly to unity.
Under pure dephasing [Fig.~\ref{fig:dynamics}(d)] the bright manifold instead empties first into the drain, after which dephasing-mediated mixing only slowly leaks dark population through a depleted bright channel; the dark manifold is still above $0.15$ at $t=30\,\hbar/\mathrm{meV}$ and the sink lags at $\eta\approx 0.8$.
This transient bright population is the distinctive signature of the one-way valve.

\paragraph*{Scaling with system size.}
The dephasing equilibration timescale $1/(\gamma_{\mathrm{deph}}\Lambda)$ degrades as the dark fraction~\cite{Khazanov2023,Pandya2022} 
grows, while the rescue rate is universal across the dark manifold.
The two mechanisms therefore produce qualitatively different scaling of peak transport efficiency with system size. Fig.~\ref{fig:schematic}(c) reports the peak efficiency for $N$ from 3 to 96, with $\gamma_{\mathrm{rec}}$ and $\gamma_{\mathrm{deph}}$ each independently optimized at every $N$, and obtained by solving full Lindbladian.
Rescue channel, $\eta^*(\gamma_{\rm rec})$ (blue) maintains essentially complete transport, $\eta^*_{\mathrm{rec}} > 0.998$, across the entire range, whereas $\eta^* (\gamma_{\rm deph})$ (orange) degrades with $N$ as its growing dark manifold throttles dephasing-mediated clearance.
As we show below in detail, the scaling of transport with system size is the smoking gun signature of an active rescue channel.
 Moreover, for optimized transport $\eta^*(\gamma_{\rm rec}, \gamma_{\rm deph})$ (pink markers, Fig.~\ref{fig:schematic}(c)) is dominated by the complete efficiency of the rescue channel.

\begin{figure}[t]
\centering
\includegraphics[width=\linewidth]{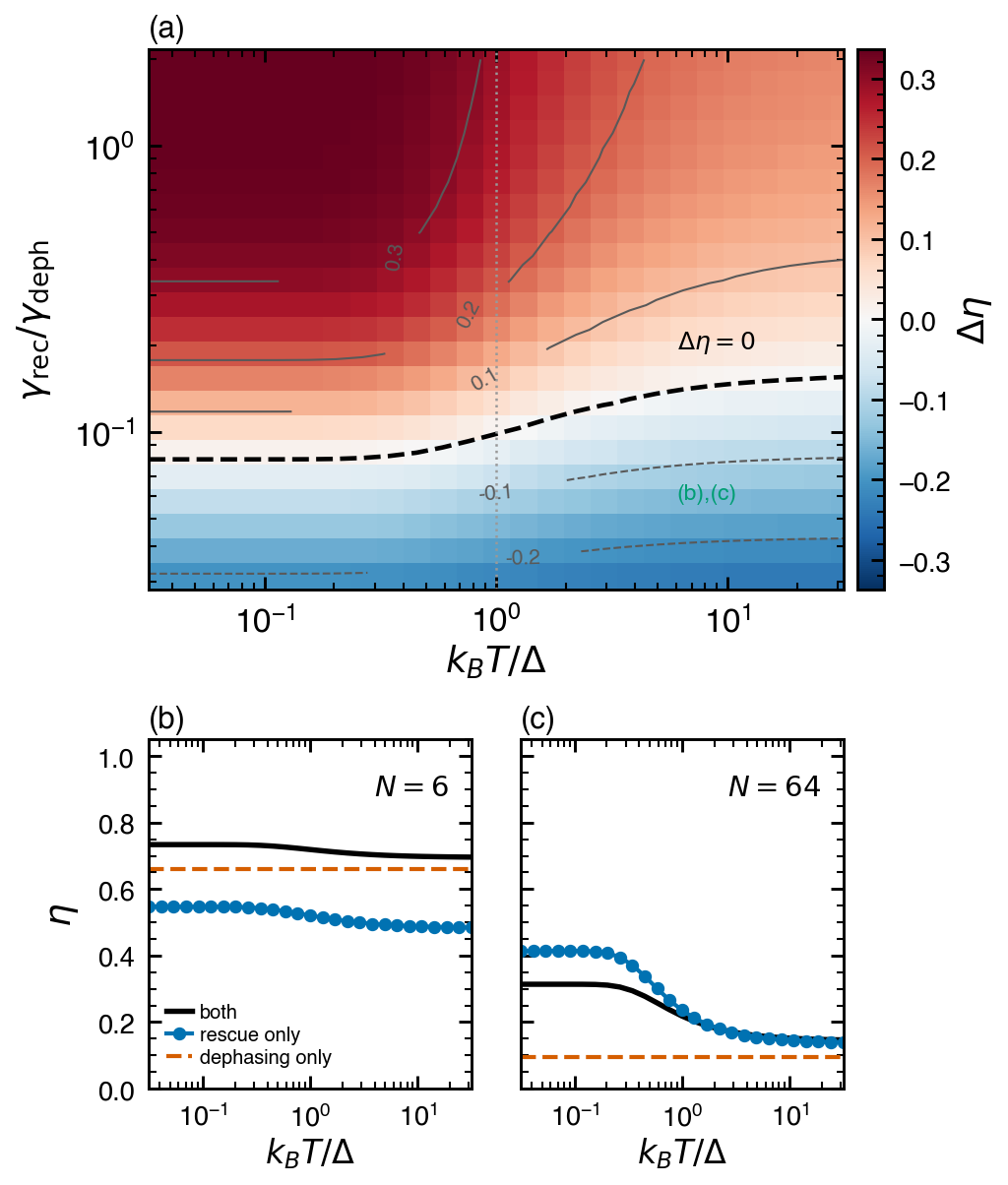}
\caption{Rescue--dephasing competition at finite temperature.
(a) Efficiency difference $\Delta\eta = \eta_{\mathrm{rec}} - \eta_{\mathrm{deph}}$ vs $k_BT/\Delta$ and rate ratio $\gamma_{\mathrm{rec}}/\gamma_{\mathrm{deph}}$ for $N=6$.
The absorption rate $\gamma_{\mathrm{abs}} = e^{-\Delta/k_BT}\gamma_{\mathrm{rec}}$ refills the dark manifold at finite $T$, degrading the unidirectional rescue. The crossover at $k_BT/\Delta \sim 1$ is a clean detailed-balance signature, and shifts the $\Delta\eta = 0$ boundary (black dashed) only weakly, from $\gamma_{\mathrm{rec}}/\gamma_{\mathrm{deph}} \approx 0.08$ to $\approx 0.16$.
(b),(c) Efficiency vs.\ temperature at $\gamma_{\mathrm{rec}} = 0.05\,\gamma_{\mathrm{deph}}$ for (b) $N=6$ and (c) $N=64$: both channels active (black), rescue only ($\gamma_{\mathrm{deph}} = 0$, blue), and dephasing only ($\gamma_{\mathrm{rec}} = 0$, orange dashed; $T$-independent). Raising $N$ maps the temperature dependence of rescue channel on total transport, an important experimental discriminator.}
\label{fig:temperature}
\end{figure}

\paragraph*{Finite-temperature competition.}
At finite temperature, the reverse (absorption) channel is 
reactivated with detailed-balance ratio $\gamma_{\mathrm{abs}}/\gamma_{\mathrm{rec}} = e^{-\Delta/k_BT}$, refilling the dark manifold and degrading the one-way valve, with a crossover at $k_BT/\Delta\sim1$ absent from purely electronic mechanisms~\cite{Lindlau2019,Brem2020}.
Fig.~\ref{fig:temperature}(a) maps the ensuing rescue-dephasing competition for $N=6$ at unoptimized rates representative of experimental linewidths: rescue prevails at every temperature once $\gamma_{\mathrm{rec}}\gtrsim0.1\,\gamma_{\mathrm{deph}}$, the $\Delta\eta=0$ boundary shifting only weakly with $T$; $\Delta\eta = \eta_{\rm rec} - \eta_{\rm deph}$. In particular, this crossover for $\gamma_{\rm rec}/\gamma_{\rm deph}$ decreases with increasing $N$; see SM for the corresponding phase diagram for $N=64$.
To show the effect of $N$ more vividly, in Fig.~\ref{fig:temperature}(b,c) we show variation of $\eta$ with $k_BT/\Delta$ for  $\gamma_{\mathrm{rec}}=0.05\,\gamma_{\mathrm{deph}}$. For $N=6$, in Fig.~\ref{fig:temperature}(b) this maps a line-cut below the $\Delta\eta = 0$ crossover, where dephasing still outperforms pure rescue. However, at $N=64$, in Fig.~\ref{fig:temperature}(c) the rescue channel dominates as the dephasing channel degrades with system size. As a result, the total transport at large $N$ follows the temperature dependence of rescue channel underscoring clear experimental signatures to probe an active rescue channel. 
The high-$T$ degradation simultaneously steepens with $N$ as absorption refills the larger dark manifold; both trends track the dark-state count and separate the rescue channel from any equilibration-based mechanism.
Additionally, the time dynamics of Fig.~\ref{fig:dynamics}(a) provide the corresponding kinetic fingerprint. Single-exponential dark-state decay versus a long-lived plateau, resolvable by transient absorption or time-resolved photoluminescence~\cite{DelPo2020,Pandya2022,Khazanov2023}. Since $\gamma_{\mathrm{rec}}=2J(\Delta)$, tuning the cavity detuning across a vibrational bath mode resonantly enhances the rescue rate, making transport a spectroscopy of the non-Condon spectral density~\cite{Seibt2018,Takahashi2022,Odewale2024,Kano2002}.

\paragraph*{Conclusion.}
In summary, we have identified non-Condon contribution of exchange between quantum system and phononic modes opens a directional channel where the total escape rate from any dark to bright state is pinned 
by the photonic-weight sum rule, independent of system size. 
Experimentally, Herzberg-Teller polaritons identified spectroscopically in metalloporphyrin microcavities~\cite{Seibt2018,Takahashi2022,Odewale2024} confirm that non-Condon coupling is operative in real materials, with $\gamma_{\mathrm{rec}}=2J(\Delta)$ estimated to reach the meV scale set by the measured Condon dephasing and hence comparable to cavity leakage (also see SM~\cite{SuppMaterial}). The rescue channel should be most accessible in solid-state platforms whose transition dipole is phonon-modulated such as organic semiconductor microcavities and molecular crystals that already display strong-coupling exciton transport~\cite{Lidzey1998,Kena-Cohen2010,Berghuis2022}, transition-metal-dichalcogenides momentum-dark excitons radiate only through phonon-assisted, Herzberg--Teller-like channels~\cite{Lindlau2019,Brem2020}, intersubband and Landau-polariton devices with strong engineered light--matter coupling~\cite{Paravicini-Bagliani2019}, nanoplatelet and quantum-dot solids with pronounced deformation-potential phonon coupling~\cite{Emin1987}, and photonic-crystal or plasmonic reservoirs with structured spectral densities~\cite{Gonzalez-Ballestero2015}. The size-scaling test (Fig.~\ref{fig:schematic}(c)) calls for $N\gtrsim 10$, where the predicted advantage exceeds $\sim 30\%$, while the temperature signature (Fig.~\ref{fig:temperature}) is cleanest at $\Delta$ of tens of meV, putting $T_{\mathrm{crit}}\!\sim\!100$~K within reach of cryogenic cavity microscopy.

For quantum array systems, several directions follow. Cold-atom cavity-QED arrays with tunable Raman couplings~\cite{Haroche1989,Raimond2001,aidelsburger2011experimental,du2016experimental} offer a clean emulator in which $\gamma_{\mathrm{rec}}$, $\gamma_{\mathrm{deph}}$, and $N$ are independently controllable, and a route to harnessing collective dark-state dissipation for open-system quantum control. Superconducting platforms add a distinct advantage with their demonstrated compatibility with hybrid material integration~\cite{wang2022hexagonal,manenti2017circuit,kitzman2022phononic,oliver2013materials}, combined with reservoir engineering~\cite{metelmann2015nonreciprocal,mundhada2018experimental,kannan2023demand,almanakly2026driven}, turn the spectral density into an engineered design parameter.
On the theory side, extending the framework beyond the single-excitation manifold to finite pumping, to finite temperature where the absorption channel reactivates, and to structured non-Markovian baths where $J(\Delta)$ can be resonantly shaped will determine how far the one-way valve can be pushed in realistic devices.
\begin{acknowledgments}
We thank John Philbin and Jugal Talukdar for useful discussions. This work is supported by the Department of Energy (DOE) Office of Science (SC) under DE-FOA-0003432 and by Grant No. GBMF12976 of the Gordon and Betty Moore Foundation. We also acknowledge support from the National Science Foundation QuSeC-TAQS Program under Grant No. 2326840.

\end{acknowledgments}

\bibliography{references}

@article{Cao2020,
  author  = {Cao, Jianshu and Cogdell, Richard J. and Coker, David F. and Duan, Hong-Guang and Hauer, Jürgen and Kleinekathöfer, Ulrich and Jansen, Thomas L. C. and Mančal, Tomáš and Miller, R. J. Dwayne and Ogilvie, Jennifer P. and Prokhorenko, Valentyn I. and Renger, Thomas and Tan, Howe-Siang and Tempelaar, Roel and Thorwart, Michael and Thyrhaug, Erling and Westenhoff, Sebastian and Zigmantas, Donatas},
  title   = {Quantum biology revisited},
  journal = {Science Advances},
  volume  = {6},
  pages   = {eaaz4888},
  year    = {2020},
  doi     = {10.1126/sciadv.aaz4888}
}

@article{Lambert2013,
  author  = {Lambert, Neill and Chen, Yueh-Nan and Cheng, Yuan-Chung and Li, Che-Ming and Chen, Guang-Yin and Nori, Franco},
  title   = {Quantum biology},
  journal = {Nature Physics},
  volume  = {9},
  pages   = {10--18},
  year    = {2013},
  doi     = {10.1038/nphys2474}
}

@article{Anderson1958,
  author  = {Anderson, Philip W.},
  title   = {Absence of Diffusion in Certain Random Lattices},
  journal = {Physical Review},
  volume  = {109},
  pages   = {1492--1505},
  year    = {1958},
  doi     = {10.1103/PhysRev.109.1492}
}

@book{Emin1987,
  author    = {Emin, David},
  title     = {Phonons in Semiconductor Nanostructures},
  publisher = {Springer},
  year      = {1987},
  note      = {Hopping transport in disordered systems}
}

@article{Schachenmayer2015,
  author  = {Schachenmayer, Johannes and Genes, Claudiu and Tignone, Edoardo and Pupillo, Guido},
  title   = {Cavity-Enhanced Transport of Excitons},
  journal = {Physical Review Letters},
  volume  = {114},
  pages   = {196403},
  year    = {2015},
  doi     = {10.1103/PhysRevLett.114.196403}
}

@article{FeistGarciaVidal2015,
  author  = {Feist, Johannes and Garcia-Vidal, Francisco J.},
  title   = {Extraordinary Exciton Conductance Induced by Strong Coupling},
  journal = {Physical Review Letters},
  volume  = {114},
  pages   = {196402},
  year    = {2015},
  doi     = {10.1103/PhysRevLett.114.196402}
}

@article{Orgiu2015,
  author  = {Orgiu, Emanuele and George, Jino and Hutchison, James A. and Devaux, Eloise and Dayen, Jean F. and Doudin, Bernard and Stellacci, Francesco and Genet, Cyriaque and Schachenmayer, Johannes and Genes, Claudiu and Pupillo, Guido and Samorì, Paolo and Ebbesen, Thomas W.},
  title   = {Conductivity in organic semiconductors hybridized with the vacuum field},
  journal = {Nature Materials},
  volume  = {14},
  pages   = {1123--1129},
  year    = {2015},
  doi     = {10.1038/nmat4392}
}

@article{Zhong2016,
  author  = {Zhong, Xiaolan and Chervy, Thibault and Wang, Shaojun and George, Jino and Thomas, Anoop and Hutchison, James A. and Devaux, Eloise and Genet, Cyriaque and Ebbesen, Thomas W.},
  title   = {Non-Radiative Energy Transfer Mediated by Hybrid Light-Matter States},
  journal = {Angewandte Chemie International Edition},
  volume  = {55},
  pages   = {6202--6206},
  year    = {2016},
  doi     = {10.1002/anie.201600428}
}

@article{Paravicini-Bagliani2019,
  author  = {Paravicini-Bagliani, Gian L. and Appugliese, Felice and Richter, Eli and Valmorra, Filippo and Keller, Jérôme and Beck, Mattias and Bartolo, Nicolas and Rössler, Clemens and Ihn, Thomas and Ensslin, Klaus and Ciuti, Cristiano and Scalari, Giacomo and Faist, Jérôme},
  title   = {Magneto-transport controlled by Landau polariton states},
  journal = {Nature Physics},
  volume  = {15},
  pages   = {186--190},
  year    = {2019},
  doi     = {10.1038/s41567-018-0346-y}
}

@article{Agranovich2003,
  author  = {Agranovich, Vladimir M. and Litinskaya, Marina and Lidzey, David G.},
  title   = {Cavity polaritons in microcavities containing disordered organic semiconductors},
  journal = {Physical Review B},
  volume  = {67},
  pages   = {085311},
  year    = {2003},
  doi     = {10.1103/PhysRevB.67.085311}
}

@article{Gonzalez-Ballestero2015,
  author  = {González-Ballestero, Carlos and Feist, Johannes and Moreno, Esteban and Garcia-Vidal, Francisco J.},
  title   = {Harvesting excitons through plasmonic strong coupling},
  journal = {Physical Review B},
  volume  = {92},
  pages   = {121402},
  year    = {2015},
  doi     = {10.1103/PhysRevB.92.121402}
}

@article{Gonzalez-Ballestero2016,
  author  = {González-Ballestero, Carlos and Feist, Johannes and Gonzalo Badía, Eduardo and Moreno, Esteban and García-Vidal, Francisco J.},
  title   = {Uncoupled Dark States Can Inherit Polaritonic Properties},
  journal = {Physical Review Letters},
  volume  = {117},
  pages   = {156402},
  year    = {2016},
  doi     = {10.1103/PhysRevLett.117.156402}
}

@article{Houdre1996,
  author  = {Houdré, R. and Stanley, R. P. and Ilegems, M.},
  title   = {Vacuum-field Rabi splitting in the presence of inhomogeneous broadening: Resolution of a homogeneous linewidth in an inhomogeneously broadened system},
  journal = {Physical Review A},
  volume  = {53},
  pages   = {2711--2715},
  year    = {1996},
  doi     = {10.1103/PhysRevA.53.2711}
}

@article{DelPo2020,
  author  = {DelPo, Christina A. and Kudisch, Bryan and Park, Kwang Hyun and Khan, Sana-Ul-Haq and Griebel, Florian and Balasubrahmaniyam, Mani and Lukens, Jesse T. and Gutiérrez, Margarita and Maiuri, Margherita and Scholes, Gregory D.},
  title   = {Polariton Transitions in Femtosecond Transient Absorption Studies of Ultrastrong Light-Molecule Coupling},
  journal = {The Journal of Physical Chemistry Letters},
  volume  = {11},
  pages   = {2667--2674},
  year    = {2020},
  doi     = {10.1021/acs.jpclett.0c00247}
}

@article{Pandya2022,
  author  = {Pandya, Raj and Ashoka, Arjun and Georgiou, Kyriacos and Sung, Jooyoung and Jayaprakash, Rahul and Renken, Scott and Gai, Lizhi and Shen, Zhen and Rao, Akshay and Musser, Andrew J.},
  title   = {Tuning the Coherent Propagation of Organic Exciton-Polaritons through Dark State Delocalization},
  journal = {Advanced Science},
  volume  = {9},
  pages   = {2105569},
  year    = {2022},
  doi     = {10.1002/advs.202105569}
}

@article{Berghuis2022,
  author  = {Berghuis, A. Marios and Halpin, Aidan and Le-Van, Quynh and Ramezani, Mohammad and Wang, Shaojun and Murai, Shunsuke and Rivas, Jaime Gómez},
  title   = {Controlling Exciton Propagation in Organic Crystals through Strong Coupling to Plasmonic Nanoparticle Arrays},
  journal = {ACS Photonics},
  volume  = {9},
  pages   = {2263--2272},
  year    = {2022},
  doi     = {10.1021/acsphotonics.2c00007}
}

@article{Khazanov2023,
  author  = {Khazanov, Thomas and Gunasekaran, Suman and George, Aleesha and Lomlu, Rana and Mukherjee, Soham and Musser, Andrew J.},
  title   = {Embrace the darkness: An experimental perspective on organic exciton-polaritons},
  journal = {Chemical Physics Reviews},
  volume  = {4},
  pages   = {041305},
  year    = {2023},
  doi     = {10.1063/5.0168948}
}

@article{Sandik2024,
  author  = {Sandik, Gal and Feist, Johannes and García-Vidal, Francisco J. and Schwartz, Tal},
  title   = {Cavity-enhanced energy transport in molecular systems},
  journal = {Nature Materials},
  year    = {2024},
  note    = {Published online 9 August 2024},
  doi     = {10.1038/s41563-024-01962-5}
}

@article{Groenhof2019,
  author  = {Groenhof, Gerrit and Climent, Clàudia and Feist, Johannes and Morozov, Dmitry and Toppari, J. Jussi},
  title   = {Tracking Polariton Relaxation with Multiscale Molecular Dynamics Simulations},
  journal = {The Journal of Physical Chemistry Letters},
  volume  = {10},
  pages   = {5476--5483},
  year    = {2019},
  doi     = {10.1021/acs.jpclett.9b02192}
}

@article{Sokolovskii2023,
  author  = {Sokolovskii, Ilia and Tichauer, Ruth H. and Morozov, Dmitry and Feist, Johannes and Groenhof, Gerrit},
  title   = {Multi-scale molecular dynamics simulations of enhanced energy transfer in organic molecules under strong coupling},
  journal = {Nature Communications},
  volume  = {14},
  pages   = {6613},
  year    = {2023},
  doi     = {10.1038/s41467-023-42067-y}
}

@article{Davidsson2023,
  author  = {Davidsson, Eric and Kowalewski, Markus},
  title   = {Simulating photodissociation reactions in cavities with polaritonic Lindblad master equations},
  journal = {Journal of Chemical Physics},
  volume  = {158},
  pages   = {234106},
  year    = {2023},
  doi     = {10.1063/5.0148615}
}

@article{Plenio2008,
  author  = {Plenio, Martin B. and Huelga, Susana F.},
  title   = {Dephasing-assisted transport: quantum networks and biomolecules},
  journal = {New Journal of Physics},
  volume  = {10},
  pages   = {113019},
  year    = {2008},
  doi     = {10.1088/1367-2630/10/11/113019}
}

@article{Rebentrost2009,
  author  = {Rebentrost, Patrick and Mohseni, Masoud and Kassal, Ivan and Lloyd, Seth and Aspuru-Guzik, Alán},
  title   = {Environment-assisted quantum transport},
  journal = {New Journal of Physics},
  volume  = {11},
  pages   = {033003},
  year    = {2009},
  doi     = {10.1088/1367-2630/11/3/033003}
}

@article{Cao2009,
  author  = {Cao, Jianshu and Silbey, Robert J.},
  title   = {Optimization of exciton trapping in energy transfer processes},
  journal = {The Journal of Physical Chemistry A},
  volume  = {113},
  pages   = {13825--13838},
  year    = {2009},
  doi     = {10.1021/jp9032589}
}

@article{Dou2015a,
  author  = {Dou, Wenjie and Subotnik, Joseph E. and Nitzan, Abraham},
  title   = {Surface hopping with a manifold of electronic states. {II}. Application to the many-body {Anderson-Holstein} model},
  journal = {Journal of Chemical Physics},
  volume  = {142},
  pages   = {084110},
  year    = {2015},
  doi     = {10.1063/1.4908034}
}

@article{Dou2015b,
  author  = {Dou, Wenjie and Nitzan, Abraham and Subotnik, Joseph E.},
  title   = {Frictional effects near a metal surface},
  journal = {Journal of Chemical Physics},
  volume  = {143},
  pages   = {054103},
  year    = {2015},
  doi     = {10.1063/1.4927237}
}

@article{Dou2018,
  author  = {Dou, Wenjie and Subotnik, Joseph E.},
  title   = {Universality of electronic friction. {II}. Equivalence of the quantum-classical Liouville equation approach with von Oppen's nonequilibrium {Green}'s function methods out of equilibrium},
  journal = {Physical Review B},
  volume  = {97},
  pages   = {064303},
  year    = {2018},
  doi     = {10.1103/PhysRevB.97.064303}
}

@article{Haroche1989,
  author  = {Haroche, Serge and Kleppner, Daniel},
  title   = {Cavity Quantum Electrodynamics},
  journal = {Physics Today},
  volume  = {42},
  pages   = {24--30},
  year    = {1989},
  doi     = {10.1063/1.881201}
}

@article{Raimond2001,
  author  = {Raimond, Jean-Michel and Brune, Michel and Haroche, Serge},
  title   = {Manipulating quantum entanglement with atoms and photons in a cavity},
  journal = {Reviews of Modern Physics},
  volume  = {73},
  pages   = {565--582},
  year    = {2001},
  doi     = {10.1103/RevModPhys.73.565}
}

@article{Lidzey1998,
  author  = {Lidzey, David G. and Bradley, David D. C. and Skolnick, Mark S. and Virgili, Tersilla and Walker, Stuart and Whittaker, David M.},
  title   = {Strong exciton-photon coupling in an organic semiconductor microcavity},
  journal = {Nature},
  volume  = {395},
  pages   = {53--55},
  year    = {1998},
  doi     = {10.1038/25692}
}

@article{Kena-Cohen2010,
  author  = {Kéna-Cohen, Stéphane and Forrest, Stephen R.},
  title   = {Room-temperature polariton lasing in an organic single-crystal microcavity},
  journal = {Nature Photonics},
  volume  = {4},
  pages   = {371--375},
  year    = {2010},
  doi     = {10.1038/nphoton.2010.86}
}

@article{Ribeiro2022,
  author  = {Ribeiro, Raphael F.},
  title   = {Multimode polariton effects on molecular energy transport and spectral fluctuations},
  journal = {Communications Chemistry},
  volume  = {5},
  pages   = {48},
  year    = {2022},
  doi     = {10.1038/s42004-022-00660-0}
}

@article{Mandal2023,
  author  = {Mandal, Arkajit and Taylor, Michael A. D. and Weight, Braden M. and Koessler, Eric R. and Li, Xinyang and Huo, Pengfei},
  title   = {Theoretical Advances in Polariton Chemistry and Molecular Cavity Quantum Electrodynamics},
  journal = {Chemical Reviews},
  volume  = {123},
  pages   = {9786--9879},
  year    = {2023},
  doi     = {10.1021/acs.chemrev.2c00855}
}

@article{Hubener2021,
  author  = {Hübener, Hannes and De Giovannini, Umberto and Schäfer, Christian and Andberger, Josefine and Ruggenthaler, Michael and Faist, Jérôme and Rubio, Angel},
  title   = {Engineering quantum materials with chiral optical cavities},
  journal = {Nature Materials},
  volume  = {20},
  pages   = {438--442},
  year    = {2021},
  doi     = {10.1038/s41563-020-00801-7}
}

@article{Michail2024,
  author  = {Michail, Evripidis and Rashidi, Kamyar and Liu, Bin and He, Guiying and Menon, Vinod M. and Sfeir, Matthew Y.},
  title   = {Addressing the Dark State Problem in Strongly Coupled Organic Exciton-Polariton Systems},
  journal = {Nano Letters},
  volume  = {24},
  pages   = {557--565},
  year    = {2024},
  doi     = {10.1021/acs.nanolett.3c02984}
}

@article{Xiang2019,
  author  = {Xiang, Bo and Ribeiro, Raphael F. and Chen, Liying and Wang, Jiaxi and Du, Matthew and Yuen-Zhou, Joel and Xiong, Wei},
  title   = {State-Selective Polariton to Dark State Relaxation Dynamics},
  journal = {The Journal of Physical Chemistry A},
  volume  = {123},
  pages   = {5918--5927},
  year    = {2019},
  doi     = {10.1021/acs.jpca.9b04601}
}

@article{Seibt2018,
  author  = {Seibt, Joachim and Mančal, Tomáš},
  title   = {Treatment of Herzberg-Teller and non-Condon effects in optical spectra with hierarchical equations of motion},
  journal = {Chemical Physics},
  volume  = {515},
  pages   = {129--140},
  year    = {2018},
  doi     = {10.1016/j.chemphys.2018.08.026}
}

@article{PerezSanchez2025,
  author  = {Pérez-Sánchez, Juan B. and Yuen-Zhou, Joel},
  title   = {Radiative pumping vs vibrational relaxation of molecular polaritons: a bosonic mapping approach},
  journal = {Nature Communications},
  volume  = {16},
  pages   = {3067},
  year    = {2025},
  doi     = {10.1038/s41467-025-58045-5}
}

@article{Odewale2024,
  author  = {Odewale, Elizabeth O. and Avramenko, Aleksandr G. and Rury, Aaron S.},
  title   = {Deciphering between enhanced light emission and absorption in multi-mode porphyrin cavity polariton samples},
  journal = {Nanophotonics},
  volume  = {13},
  number  = {14},
  pages   = {2695--2706},
  year    = {2024},
  doi     = {10.1515/nanoph-2023-0748}
}

@article{Takahashi2022,
  author  = {Takahashi, Shunsuke and Watanabe, Kazuyuki and Matsumoto, Yoshiyasu},
  title   = {Light Emission from Vibronic Polaritons in Coupled Metalloporphyrin-Multimode Cavity Systems},
  journal = {The Journal of Physical Chemistry Letters},
  volume  = {13},
  pages   = {4187--4194},
  year    = {2022},
  doi     = {10.1021/acs.jpclett.2c00353}
}

@article{Lindlau2019,
  author  = {Lindlau, Jessica and Selig, Malte and Neumann, Alexander and Colombier, Léo and Förste, Jonathan and Funk, Victor and Förg, Michael and Kim, Jonghwan and Berghäuser, Gunnar and Taniguchi, Takashi and Watanabe, Kenji and Wang, Feng and Malic, Ermin and Högele, Alexander},
  title   = {The role of momentum-dark excitons in the elementary optical response of bilayer WSe$_2$},
  journal = {Nature Communications},
  volume  = {9},
  pages   = {2586},
  year    = {2018},
  doi     = {10.1038/s41467-018-04877-3}
}

@article{Brem2020,
  author  = {Brem, Samuel and Ekman, August and Christiansen, Dominik and Katsch, Florian and Selig, Malte and Robert, Cedric and Marie, Xavier and Urbaszek, Bernhard and Knorr, Andreas and Malic, Ermin},
  title   = {Phonon-Assisted Photoluminescence from Indirect Excitons in Monolayers of Transition-Metal Dichalcogenides},
  journal = {Nano Letters},
  volume  = {20},
  pages   = {2849--2856},
  year    = {2020},
  doi     = {10.1021/acs.nanolett.0c00633}
}

@article{Majumdar2011,
  author  = {Majumdar, Arka and Bajcsy, Michal and Rundquist, Armand and Vučković, Jelena},
  title   = {Phonon-mediated coupling between quantum dots and a photonic-crystal nanocavity},
  journal = {Physical Review B},
  volume  = {84},
  pages   = {085309},
  year    = {2011},
  doi     = {10.1103/PhysRevB.84.085309}
}

@article{RoyHughes2011,
  author  = {Roy, Chiranjeeb and Hughes, Stephen},
  title   = {Influence of Electron-Acoustic-Phonon Scattering on Intensity Power Broadening in a Coherently Driven Quantum-Dot-Cavity System},
  journal = {Physical Review X},
  volume  = {1},
  pages   = {021009},
  year    = {2011},
  doi     = {10.1103/PhysRevX.1.021009}
}

@article{Hutchison2012,
  author  = {Hutchison, James A. and Schwartz, Tal and Genet, Cyriaque and Devaux, Eloise and Ebbesen, Thomas W.},
  title   = {Modifying Chemical Landscapes by Coupling to Vacuum Fields},
  journal = {Angewandte Chemie International Edition},
  volume  = {51},
  pages   = {1592--1596},
  year    = {2012},
  doi     = {10.1002/anie.201107033}
}

@misc{SuppMaterial,
  note = {See Supplemental Material at [URL] for the rigorous Born-Markov derivation, the full Redfield-tensor treatment of the non-Condon channel, the regime of validity of the site-basis reduction, the site-$N$ drain phase diagram, and the QCLE validation.}
}

@article{Kano2002,
  author  = {Kano, Hideaki and Saito, Takashi and Kobayashi, Takayoshi},
  title   = {Observation of Herzberg-Teller-type wave packet motion in porphyrin J-aggregates studied by sub-5-fs spectroscopy},
  journal = {The Journal of Physical Chemistry A},
  volume  = {106},
  pages   = {3445--3453},
  year    = {2002},
  doi     = {10.1021/jp012493f}
}

@book{BreuerPetruccione,
  author    = {Breuer, Heinz-Peter and Petruccione, Francesco},
  title     = {The Theory of Open Quantum Systems},
  publisher = {Oxford University Press},
  year      = {2002}
}

@article{Flick18,
url = {https://doi.org/10.1515/nanoph-2018-0067},
title = {Strong light-matter coupling in quantum chemistry and quantum photonics},
author = {Johannes Flick and Nicholas Rivera and Prineha Narang},
pages = {1479--1501},
volume = {7},
number = {9},
journal = {Nanophotonics},
doi = {doi:10.1515/nanoph-2018-0067},
year = {2018},
lastchecked = {2023-07-25}
}

@article{RibeiroChemSci2018,
author = {Ribeiro, Raphael F and Mart{\'{i}}nez-Mart{\'{i}}nez, Luis A and Du, Matthew and Campos-Gonzalez-Angulo, Jorge and Yuen-Zhou, Joel},
doi = {10.1039/C8SC01043A},
journal = {Chem. Sci.},
number = {30},
pages = {6325--6339},
publisher = {The Royal Society of Chemistry},
title = {{Polariton chemistry: controlling molecular dynamics with optical cavities}},
volume = {9},
year = {2018}
}

@article{Kena-Cohen2019,
author = {K{\'{e}}na-Cohen, St{\'{e}}phane and Yuen-Zhou, Joel},
doi = {10.1021/acscentsci.9b00219},
issn = {23747951},
journal = {ACS Cent. Sci.},
number = {3},
pages = {386--388},
title = {{Polariton Chemistry: Action in the Dark}},
volume = {5},
year = {2019}
}

@article{DuChemSci2018,
author = {Du, Matthew and Mart{\'{i}}nez-Mart{\'{i}}nez, Luis A and Ribeiro, Raphael F and Hu, Zixuan and Menon, Vinod M and Yuen-Zhou, Joel},
doi = {10.1039/C8SC00171E},
journal = {Chem. Sci.},
number = {32},
pages = {6659--6669},
publisher = {The Royal Society of Chemistry},
title = {{Theory for polariton-assisted remote energy transfer}},
volume = {9},
year = {2018}
}

@article{Blazquez18,
  title = {Organic polaritons enable local vibrations to drive long-range energy transfer},
  author = {S\'aez-Bl\'azquez, R. and Feist, J. and Fern\'andez-Dom\'{\i}nguez, A. I. and Garc\'{\i}a-Vidal, F. J.},
  journal = {Phys. Rev. B},
  volume = {97},
  issue = {24},
  pages = {241407},
  numpages = {5},
  year = {2018},
  month = {Jun},
  publisher = {American Physical Society},
  doi = {10.1103/PhysRevB.97.241407},
  url = {https://link.aps.org/doi/10.1103/PhysRevB.97.241407}
}

@article{bittorf2026long,
  title={Long-Range Exciton Energy Transfer in Two-Dimensional Materials},
  author={Bittorf, Paul H and Black, Maximilian and Crispin, Hebrew and Singh, Prabhdeep and Darman, Parsa and Darbari, Sara and Chahshouri, Fatemeh and Taleb, Masoud and Talebi, Nahid},
  journal={Laser \& Photonics Reviews},
  volume={20},
  number={8},
  pages={e01604},
  year={2026},
  publisher={Wiley Online Library}
}

@article{herrera2025moire,
  title={Moir{\'e} excitons and exciton--polaritons: a review},
  author={Herrera-Gonz{\'a}lez, Sa{\'u}l A and Lara-Garc{\'\i}a, Hugo A and Pirruccio, Giuseppe and Ruiz-Tijerina, David A and Camacho-Guardian, Arturo},
  journal={Journal of Physics: Condensed Matter},
  volume={37},
  number={48},
  pages={483002},
  year={2025},
  publisher={IOP Publishing}
}

@article{low2017polaritons,
  title={Polaritons in layered two-dimensional materials},
  author={Low, Tony and Chaves, Andrey and Caldwell, Joshua D and Kumar, Anshuman and Fang, Nicholas X and Avouris, Phaedon and Heinz, Tony F and Guinea, Francisco and Martin-Moreno, Luis and Koppens, Frank},
  journal={Nature materials},
  volume={16},
  number={2},
  pages={182--194},
  year={2017},
  publisher={Nature Publishing Group UK London}
}

@article{rubies2022photon,
  title={Photon control and coherent interactions via lattice dark states in atomic arrays},
  author={Rubies-Bigorda, Oriol and Walther, Valentin and Patti, Taylor L and Yelin, Susanne F},
  journal={Physical Review Research},
  volume={4},
  number={1},
  pages={013110},
  year={2022},
  publisher={APS}
}

@article{wang2022hexagonal,
  title={Hexagonal boron nitride as a low-loss dielectric for superconducting quantum circuits and qubits},
  author={Wang, Joel IJ and Yamoah, Megan A and Li, Qing and Karamlou, Amir H and Dinh, Thao and Kannan, Bharath and Braum{\"u}ller, Jochen and Kim, David and Melville, Alexander J and Muschinske, Sarah E and others},
  journal={Nature materials},
  volume={21},
  number={4},
  pages={398--403},
  year={2022},
  publisher={Nature Publishing Group UK London}
}

@article{metelmann2015nonreciprocal,
  title={Nonreciprocal photon transmission and amplification via reservoir engineering},
  author={Metelmann, Anja and Clerk, Aashish A},
  journal={Physical Review X},
  volume={5},
  number={2},
  pages={021025},
  year={2015},
  publisher={APS}
}

@article{manenti2017circuit,
  title={Circuit quantum acoustodynamics with surface acoustic waves},
  author={Manenti, Riccardo and Kockum, Anton F and Patterson, Andrew and Behrle, Tanja and Rahamim, Joseph and Tancredi, Giovanna and Nori, Franco and Leek, Peter J},
  journal={Nature communications},
  volume={8},
  number={1},
  pages={975},
  year={2017},
  publisher={Nature Publishing Group UK London}
}

@article{kitzman2022phononic,
  title={Phononic bath engineering of a superconducting qubit},
  author={Kitzman, Joe M and Lane, Justin R and Undershute, Camryn and Harrington, PM and Beysengulov, NR and Mikolas, CA and Murch, KW and Pollanen, Johannes},
  journal={arXiv preprint arXiv:2208.07423},
  year={2022}
}

@article{aidelsburger2011experimental,
  title={Experimental realization of strong effective magnetic fields in an optical lattice},
  author={Aidelsburger, Monika and Atala, Marcos and Nascimbene, Sylvain and Trotzky, Stefan and Chen, Yu-Ao and Bloch, Immanuel},
  journal={arXiv preprint arXiv:1110.5314},
  year={2011}
}

@article{du2016experimental,
  title={Experimental realization of stimulated Raman shortcut-to-adiabatic passage with cold atoms},
  author={Du, Yan-Xiong and Liang, Zhen-Tao and Li, Yi-Chao and Yue, Xian-Xian and Lv, Qing-Xian and Huang, Wei and Chen, Xi and Yan, Hui and Zhu, Shi-Liang},
  journal={Nature communications},
  volume={7},
  number={1},
  pages={12479},
  year={2016},
  publisher={Nature Publishing Group UK London}
}

@article{mundhada2018experimental,
  title={Experimental implementation of a Raman-assisted six-quanta process},
  author={Mundhada, Shantanu O and Grimm, Alexander and Venkatraman, Jayameenakshi and Minev, Zlatko K and Touzard, Steven and Frattini, Nicholas E and Sivak, Volodymyr V and Sliwa, Katrina and Reinhold, Philip and Shankar, Shyam and others},
  journal={arXiv preprint arXiv:1811.06589},
  year={2018}
}

@article{kannan2023demand,
  title={On-demand directional microwave photon emission using waveguide quantum electrodynamics},
  author={Kannan, Bharath and Almanakly, Aziza and Sung, Youngkyu and Di Paolo, Agustin and Rower, David A and Braum{\"u}ller, Jochen and Melville, Alexander and Niedzielski, Bethany M and Karamlou, Amir and Serniak, Kyle and others},
  journal={Nature Physics},
  volume={19},
  number={3},
  pages={394--400},
  year={2023},
  publisher={Nature Publishing Group UK London}
}

@article{oliver2013materials,
  title={Materials in superconducting quantum bits},
  author={Oliver, William D and Welander, Paul B},
  journal={MRS bulletin},
  volume={38},
  number={10},
  pages={816--825},
  year={2013},
  publisher={Cambridge University Press}
}

@article{almanakly2026driven,
  title={Driven-dissipative entanglement of distant giant atoms},
  author={Almanakly, Aziza and Soro, Ariadna and Vivas-Via{\~n}a, Alejandro and Yankelevich, Beatriz and Groiseau, Caspar and Pahl, David and An, Junyoung and Cutter, Gabriel and Gingras, Michael E and Niedzielski, Bethany M and others},
  journal={arXiv preprint arXiv:2606.13375},
  year={2026}
  }

@article{flick2019excited,
  title={Excited-state nanophotonic and polaritonic chemistry with ab initio potential-energy surfaces},
  author={Flick, Johannes and Narang, Prineha},
  journal={arXiv preprint arXiv:1907.04646},
  year={2019}
}

@article{flick2018cavity,
  title={Cavity-correlated electron-nuclear dynamics from first principles},
  author={Flick, Johannes and Narang, Prineha},
  journal={Physical review letters},
  volume={121},
  number={11},
  pages={113002},
  year={2018},
  publisher={APS}
}

@article{wang2021light,
  title={Light--matter interaction of a molecule in a dissipative cavity from first principles},
  author={Wang, Derek S and Neuman, Tom{\'a}{\v{s}} and Flick, Johannes and Narang, Prineha},
  journal={The Journal of Chemical Physics},
  volume={154},
  number={10},
  year={2021},
  publisher={AIP Publishing}
}

@article{ashida2021cavity,
  title={Cavity quantum electrodynamics at arbitrary light-matter coupling strengths},
  author={Ashida, Yuto and {\.I}mamo{\u{g}}lu, Ata{\c{c}} and Demler, Eugene},
  journal={Physical Review Letters},
  volume={126},
  number={15},
  pages={153603},
  year={2021},
  publisher={APS}
}

\appendix
\onecolumngrid
\newpage

\pagebreak
\widetext

\setcounter{equation}{0}
\setcounter{figure}{0}
\setcounter{table}{0}
\setcounter{page}{1}
\makeatletter
\renewcommand{\theequation}{S\arabic{equation}}
\renewcommand{\thefigure}{S\arabic{figure}}
\renewcommand{\bibnumfmt}[1]{[S#1]}

\section*{Supplemental Material for "Unidirectional Dark-to-Bright Rescue in Cavity-Coupled Quantum Transport"}

\section{Microscopic origin of the rescue operator}
\label{sec:micro}

We derive the phenomenological Lindblad operator $\hat{L}_{\mathrm{rec},i} = \sqrt{\gamma_{\mathrm{rec}}}\,|cav\rangle\langle i|$ of the main text from a microscopic exciton-phonon Hamiltonian.
The starting point is the Tavis-Cummings system [Eq.~(1) of the main text] coupled to a local phonon bath at each site, $\hat{H}_B = \sum_{i,k}\omega_{ik}\hat{b}^\dagger_{ik}\hat{b}_{ik}$.
Linear expansion of the system parameters in the nuclear coordinates $\hat{Q}_{ik} = (\hat{b}_{ik}+\hat{b}^\dagger_{ik})/\sqrt{2}$ separates the system-bath coupling into two independent terms:
\begin{equation}
\hat{H}_{SB} = \underbrace{\sum_{i,k}\lambda^{(d)}_{ik}\hat{\sigma}^\dagger_i\hat{\sigma}_i\hat{Q}_{ik}}_{\hat{H}^{(\mathrm{C})}_{SB} \text{ (Condon)}}
\;+\;
\underbrace{\sum_{i,k}\lambda^{(o)}_{ik}(\hat{a}^\dagger\hat{\sigma}_i + \hat{\sigma}^\dagger_i\hat{a})\hat{Q}_{ik}}_{\hat{H}^{(\mathrm{NC})}_{SB} \text{ (non-Condon)}}.
\label{eq:HSB}
\end{equation}
The Condon term is the standard Holstein coupling that modulates site energies; the non-Condon term modulates the cavity-emitter transition dipole and is the leading correction beyond the Condon approximation.

\paragraph{Rotating-wave approximation.}
Expanding $\hat{H}^{(\mathrm{NC})}_{SB}$ produces four terms in the interaction picture: $\hat{a}^\dagger\hat{\sigma}_i\hat{b}^\dagger_{ik}$, $\hat{a}^\dagger\hat{\sigma}_i\hat{b}_{ik}$, and their Hermitian conjugates.
For $\Delta = \varepsilon - \omega_a > 0$ (emitter above cavity) and $\Delta, \omega_{ik} \gg \lambda^{(o)}, g$, the rotating-wave approximation retains only the energy-conserving emission (and absorption) processes:
\begin{equation}
\hat{H}^{(\mathrm{NC,RWA})}_{SB} = \frac{1}{\sqrt{2}}\sum_{i,k}\lambda^{(o)}_{ik}\bigl(\hat{a}^\dagger\hat{\sigma}_i\,\hat{b}^\dagger_{ik} + \hat{\sigma}^\dagger_i\hat{a}\,\hat{b}_{ik}\bigr).
\label{eq:HSB_RWA}
\end{equation}

\paragraph{Born-Markov reduction.}
Tracing out the thermal bath in the standard second-order Born-Markov framework yields the dissipator
\begin{equation}
\mathcal{D}^{(\mathrm{NC})}[\hat{\rho}] = \sum_i\Bigl\{\gamma^{(\mathrm{em})}_i\,\mathcal{L}\bigl[\hat{a}^\dagger\hat{\sigma}_i\bigr]\hat{\rho} + \gamma^{(\mathrm{abs})}_i\,\mathcal{L}\bigl[\hat{\sigma}^\dagger_i\hat{a}\bigr]\hat{\rho}\Bigr\},
\label{eq:dissipator}
\end{equation}
with $\mathcal{L}[\hat{O}]\hat{\rho} = \hat{O}\hat{\rho}\hat{O}^\dagger - \tfrac{1}{2}\{\hat{O}^\dagger\hat{O},\hat{\rho}\}$ and rates set by the non-Condon spectral density $J(\omega) = (\pi/2)\sum_k|\lambda^{(o)}_k|^2\delta(\omega - \omega_k)$ at the site-cavity detuning:
\begin{equation}
\gamma^{(\mathrm{em})}(\Delta,T) = 2J(\Delta)[1+n(\Delta,T)], \qquad
\gamma^{(\mathrm{abs})}(\Delta,T) = 2J(\Delta)\,n(\Delta,T),
\label{eq:rates}
\end{equation}
where $n(\omega,T) = [\exp(\omega/k_BT) - 1]^{-1}$.
In the single-excitation manifold $\hat{a}^\dagger\hat{\sigma}_i \to |cav\rangle\langle i|$, identifying the emission term with the rescue operator of the main text [Eq.~(4)]:
\begin{equation}
\hat{L}_{\mathrm{rec},i} = \sqrt{\gamma^{(\mathrm{em})}}\,|cav\rangle\langle i|, \qquad \hat{L}_{\mathrm{abs},i} = \sqrt{\gamma^{(\mathrm{abs})}}\,|i\rangle\langle cav|.
\label{eq:identification}
\end{equation}
The corresponding Born-Markov treatment of $\hat{H}^{(\mathrm{C})}_{SB}$ produces the dephasing dissipator $\mathcal{D}^{(\mathrm{C})}[\hat{\rho}] = \sum_i \gamma_{\mathrm{deph}}\mathcal{L}[|i\rangle\langle i|]\hat{\rho}$ with $\gamma_{\mathrm{deph}} = 2J^{(d)}(0)$ from the Condon spectral density, recovering the standard ENAQT framework~\cite{Plenio2008,Xiang2019}.

The two channels are independent terms of the underlying exciton-vibration coupling.
The relative strength $|\lambda^{(o)}/\lambda^{(d)}|^2$ is generically of order unity in molecular systems with intensity-borrowing or weakly allowed transitions, the regime targeted in this work: porphyrin J-aggregates, metalloporphyrin microcavities, and chlorophyll~\cite{Kano2002,Takahashi2022,Odewale2024}.
This ratio also fixes the absolute scale of the rescue rate. Since the two channels share the same bath and differ only through their coupling prefactors, $\gamma_{\mathrm{rec}}/\gamma_{\mathrm{deph}} = J^{(o)}(\Delta)/J^{(d)}(0) \sim |\lambda^{(o)}/\lambda^{(d)}|^2 \sim \mathcal{O}(1)$ in this regime.
Condon pure-dephasing rates in molecular polariton systems are established to be of order meV, so $\gamma_{\mathrm{rec}} = 2J(\Delta)$ is likewise of order meV, comparable to the cavity leakage $\gamma_{\mathrm{lead}}$ set by photon lifetimes of a few picoseconds ($\hbar/\tau \sim 0.1$--$1$~meV).
The rescue channel therefore operates on the same timescale as photon extraction, which is precisely the regime in which the one-way valve delivers near-unity finite-time transport.
The ratio is small for the bright K-valley exciton manifold of TMD monolayers that dominates standard cavity-polariton experiments, where deformation-potential coupling to LA/LO phonons is diagonal and modulates the transition energy rather than the transition dipole.
The notable TMD exception is phonon-assisted emission from momentum- and spin-dark excitons in W-based monolayers and bilayers, where the zero-phonon transition is forbidden and the optically active channel is intrinsically Herzberg-Teller-like~\cite{Lindlau2019,Brem2020}.
Extending the present framework to that regime would require a multi-valley Hamiltonian with explicit chiral phonon modes and is left for future work.

\section{Rigorous Redfield treatment and regime of validity}
\label{sec:redfield}

The Born-Markov derivation in the previous section invokes the secular approximation in the site basis.
Working instead in the eigenbasis of $\hat{H}$ with eigenstates $\{|\psi_k\rangle\}$ and eigenvalues $\{E_k\}$, the same procedure produces the full Redfield tensor
\begin{equation}
\mathcal{D}^{(\mathrm{NC})}[\hat{\rho}] = \sum_{i,k,l}\Gamma^{\mathrm{em}}_i(\omega_{lk})\,
w_k\,\bigl|\langle i|\psi_l\rangle\bigr|^2\,
\mathcal{L}\bigl[|\psi_k\rangle\langle\psi_l|\bigr]\hat{\rho},
\label{eq:redfield}
\end{equation}
with $\omega_{lk} = E_l - E_k$ and $\Gamma^{\mathrm{em}}_i(\omega) = 2J_i(\omega)[1+n(\omega)]$.

Equation~\eqref{eq:redfield} contains the cavity-weight prefactor $w_k$ on every transition.
For dark targets ($k\in\mathcal{D}$, $w_k = 0$), the rate vanishes identically, regardless of $J_i(\omega)$ and regardless of $T$.
\textbf{The unidirectional dark-to-bright valve property of the main text is therefore a property of the Redfield tensor itself, not just of the secular site-basis reduction.}

\paragraph{Reduction to the site-basis Lindbladian.}
For a non-Condon spectral density that is approximately flat over the range of eigenstate splittings, $J_i(\omega_{lk}) \approx J_i(\Delta)$, the Redfield sum collapses to the site-basis form via the operator identity (see e.g.~\cite{BreuerPetruccione})
\begin{equation}
\sum_{k,l}\langle\psi_k|cav\rangle\langle i|\psi_l\rangle\,|\psi_k\rangle\langle\psi_l| = |cav\rangle\langle i|,
\end{equation}
recovering Eq.~\eqref{eq:identification}.
The condition for validity of this reduction is that $J(\omega)$ vary slowly over the polariton splitting scale, i.e.,
\begin{equation}
\frac{J'(\Delta)\,\sqrt{N}\,g}{J(\Delta)} \ll 1.
\label{eq:validity}
\end{equation}
For a smooth ohmic or super-ohmic bath with cutoff $\omega_c \gtrsim \sqrt{N}g$, condition~\eqref{eq:validity} is satisfied.
For molecular polariton systems with intramolecular vibrational frequencies $\sim 100$ meV and polariton splittings of a few meV, the approximation is comfortably accurate.
For our main-text parameters ($g = 1.5$ meV, $N = 6$, so $\sqrt{N}g \approx 3.7$ meV), the approximation requires bath cutoff $\omega_c \gtrsim 5$ meV, marginal but satisfied for typical phonon spectra.
Corrections away from this limit are controlled by the spectral slope $J'(\Delta)$ and introduce $\mathcal{O}(30\%)$ modifications to $\gamma_{\mathrm{rec}}$ in the worst case.

\section{Combined Condon-non-Condon dynamics}
\label{sec:combined}

When both channels are simultaneously active, the total Lindblad equation reads
\begin{equation}
\dot{\hat{\rho}} = -i[\hat{H},\hat{\rho}] + \mathcal{D}^{(\mathrm{rec})}[\hat{\rho}] + \mathcal{D}^{(\mathrm{deph})}[\hat{\rho}] + \mathcal{D}^{(\mathrm{lead})}[\hat{\rho}].
\end{equation}
Decomposing the density matrix into bright, dark, and sink populations $(p_B, p_D, p_S)$ and applying the secular approximation yields the lumped rate equations
\begin{align}
\dot{p}_B &= -\Gamma_B p_B + \gamma_{\mathrm{rec}}\,p_D + \gamma_{\mathrm{deph}}\Lambda(p_D - p_B), \label{eq:lumped_B}\\
\dot{p}_D &= -\Gamma_D p_D - \gamma_{\mathrm{rec}}\,p_D - \gamma_{\mathrm{deph}}\Lambda(p_D - p_B), \label{eq:lumped_D}\\
\dot{p}_S &= \Gamma_B p_B + \Gamma_D p_D, \label{eq:lumped_S}
\end{align}
with $\Gamma_B = \gamma_{\mathrm{lead}}\sum_{k\in\mathcal{B}}|\langle\mathcal{X}|\psi_k\rangle|^2$ and $\Gamma_D$ analogous, where $|\mathcal{X}\rangle = |N\rangle$ for the site-$N$ drain and $|\mathcal{X}\rangle = |cav\rangle$ for the cavity drain.
The geometric overlap factor is $\Lambda = \sum_{k\in\mathcal{B},l\in\mathcal{D}}\sum_i|\langle\psi_k|i\rangle|^2|\langle i|\psi_l\rangle|^2$.

\paragraph{Steady-state dark fraction.}
In the drain-free limit ($\Gamma_B,\Gamma_D \to 0$), setting $\dot{p}_D = 0$ in Eq.~\eqref{eq:lumped_D} gives the quasi-stationary ratio
\begin{equation}
\frac{p_D^{\mathrm{qs}}}{p_B^{\mathrm{qs}}} = \frac{\gamma_{\mathrm{deph}}\Lambda}{\gamma_{\mathrm{rec}} + \gamma_{\mathrm{deph}}\Lambda} = \frac{1}{1 + \gamma_{\mathrm{rec}}/(\gamma_{\mathrm{deph}}\Lambda)}.
\label{eq:dark_fraction}
\end{equation}
In the pure-rescue limit ($\gamma_{\mathrm{deph}} \to 0$), the dark fraction approaches zero.
In the pure-dephasing limit ($\gamma_{\mathrm{rec}} \to 0$), it approaches the density-of-states-weighted equilibrium value (unity for degenerate manifolds).
Equation~\eqref{eq:dark_fraction} is the analytical statement of the structural advantage of rescue at the level of the steady state.

\paragraph{Closed-form transport efficiency.}
In the pure-rescue limit ($\gamma_{\mathrm{deph}} = 0$), Eqs.~\eqref{eq:lumped_B}-\eqref{eq:lumped_S} admit an exact solution.
With initial conditions $p_B(0) = p_B^0$, $p_D(0) = p_D^0$ (with $p_B^0 + p_D^0 = 1$), and $p_S(0) = 0$,
\begin{align}
p_D(t) &= p_D^0\,e^{-\alpha t}, \\
p_B(t) &= \left(p_B^0 + \frac{\gamma_{\mathrm{rec}}\,p_D^0}{\alpha - \Gamma_B}\right)e^{-\Gamma_B t} - \frac{\gamma_{\mathrm{rec}}\,p_D^0}{\alpha - \Gamma_B}\,e^{-\alpha t},
\end{align}
with $\alpha \equiv \Gamma_D + \gamma_{\mathrm{rec}}$.
The cumulative drain efficiency at measurement time $T$ is
\begin{equation}
\eta(T) = 1 - A\,e^{-\Gamma_B T} - B\,e^{-\alpha T},
\label{eq:closed_form_eta}
\end{equation}
with $A = p_B^0 + \gamma_{\mathrm{rec}}\,p_D^0/(\alpha - \Gamma_B)$, $B = (\Gamma_D - \Gamma_B)\,p_D^0/(\alpha - \Gamma_B)$, and $A + B = 1$.
In the cavity-drain configuration ($\Gamma_D = 0$, $F_D = 0$) the dark channel decays at $\alpha = \gamma_{\mathrm{rec}}$ and transport is limited only by the slower of $\Gamma_B^{-1}$ and $\gamma_{\mathrm{rec}}^{-1}$.
Equation~\eqref{eq:closed_form_eta} provides a direct two-exponential fit form for time-resolved cavity-emission measurements. The fast rise determines $\Gamma_B$ and the slow tail determines $\gamma_{\mathrm{rec}}$, allowing the microscopic rescue rate of Eq.~\eqref{eq:rates} to be extracted from kinetic data.

\begin{figure}[h]
\centering
\includegraphics[width=0.95\linewidth]{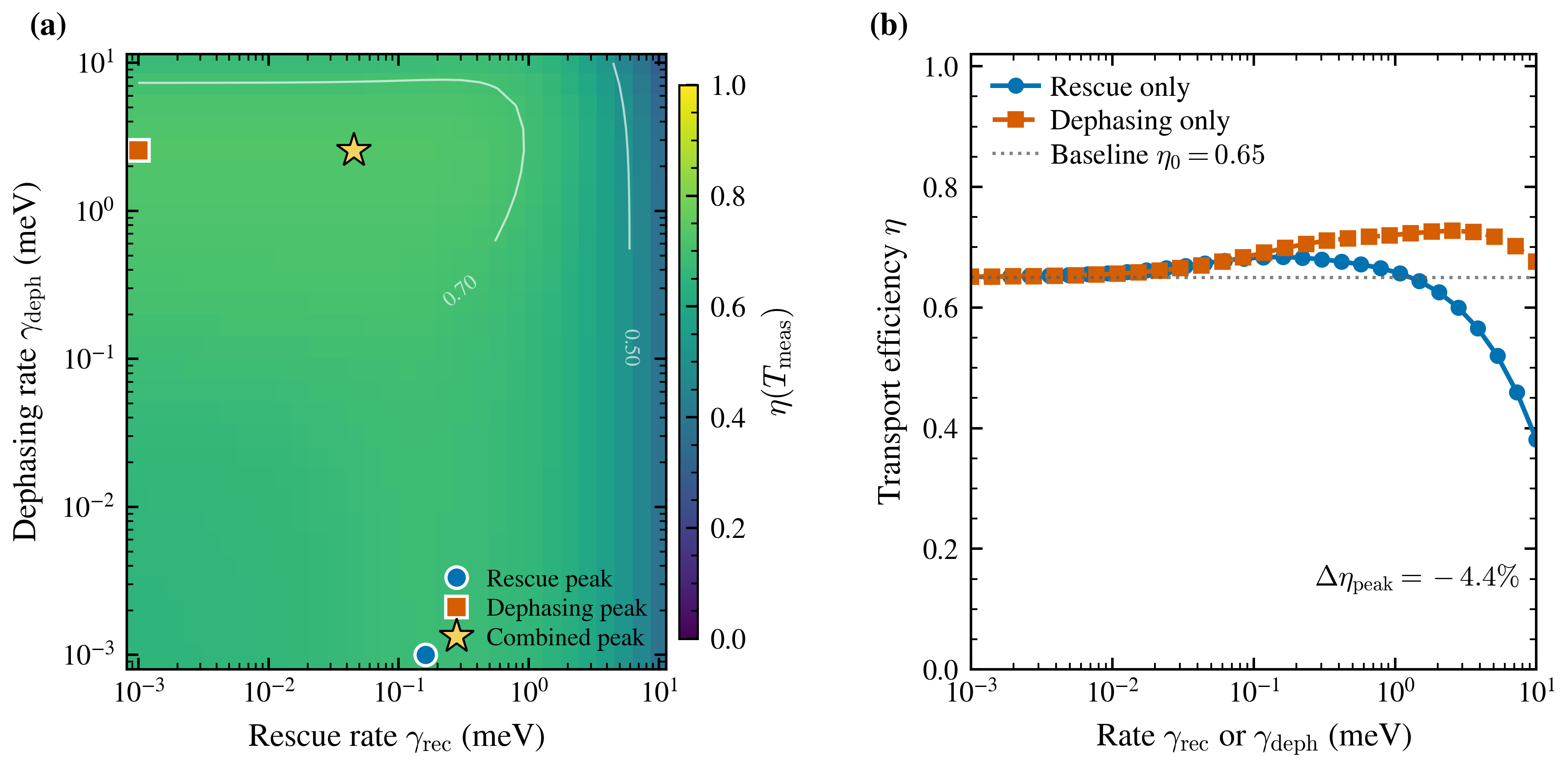}
\caption{Site-$N$ drain phase diagram, $N=6$, $g=1.5$ meV, $\delta t=0.5$ meV, $T=25\,\hbar$/meV. (a) Transport efficiency $\eta(\gamma_{\mathrm{rec}},\gamma_{\mathrm{deph}})$. (b) Pure-channel cuts: dephasing slightly outperforms rescue in this configuration ($\Delta\eta_{\mathrm{peak}} = -4.4\%$). The cavity-drain configuration of the main text reverses this verdict.}
\label{fig:siteN_phase_diagram}
\end{figure}

\paragraph{Site-$N$ drain configuration.}
For the site-$N$ drain configuration, the dark manifold has direct access to the drain ($\Gamma_D \sim \gamma_{\mathrm{lead}}$), so the steady-state dark-fraction asymmetry of Eq.~\eqref{eq:dark_fraction} does not bottleneck finite-time transport.
Pure dephasing performs the standard ENAQT delocalization function and delivers comparable transport efficiency to pure rescue at the parameters of the main text.
Figure~\ref{fig:siteN_phase_diagram} shows the numerical $\eta(\gamma_{\mathrm{rec}},\gamma_{\mathrm{deph}})$ phase diagram for this configuration.
At the optimum, dephasing slightly exceeds rescue ($\Delta\eta_{\mathrm{peak}} = -4.4\%$).

\paragraph{Regime sweep.}
Table~\ref{tab:regime} summarizes the verdict across seven parameter regimes.
The rescue advantage emerges only when the drain has zero overlap with the dark manifold ($\Gamma_D = 0$), realized physically by the cavity-drain configuration.

\begin{table}[h]
\centering
\small
\caption{Pure-channel peak transport efficiencies across parameter regimes. ``Drain'' indicates which state the drain operator extracts from. The rescue advantage emerges in the cavity-drain configurations.}
\label{tab:regime}
\begin{tabular}{lccccccc}
\toprule
Scenario & $N$ & $g$ (meV) & $T$ ($\hbar$/meV) & Drain & $\eta^*_{\mathrm{rec}}$ & $\eta^*_{\mathrm{deph}}$ & $\Delta\eta_{\mathrm{peak}}$\\
\midrule
Site-$N$ baseline           & 6  & 1.5 & 25  & site $N$ & 0.684 & 0.727 & $-4.4\%$ \\
Longer time                 & 6  & 1.5 & 100 & site $N$ & 0.990 & 0.995 & $-0.5\%$ \\
Larger system               & 10 & 1.5 & 25  & site $N$ & 0.468 & 0.520 & $-5.3\%$ \\
Stronger coupling           & 6  & 3.0 & 25  & site $N$ & 0.664 & 0.771 & $-10.7\%$\\
\textbf{Cavity drain}       & 6  & 1.5 & 25  & cav      & \textbf{1.000} & \textbf{0.799} & \textbf{$+20.1\%$}\\
\textbf{Cavity drain, large $N$} & 10 & 1.5 & 25  & cav      & \textbf{1.000} & \textbf{0.645} & \textbf{$+35.5\%$}\\
Cavity drain, longer time   & 6  & 1.5 & 100 & cav      & 1.000 & 0.998 & $+0.2\%$ \\
\bottomrule
\end{tabular}
\end{table}

\paragraph{Robustness to coherent detuning.}
The main-text transport simulations evaluate the coherent Hamiltonian at cavity resonance ($\varepsilon = \omega_a$); the site--cavity detuning $\Delta$ enters only through the rescue rate $\gamma_{\mathrm{rec}} = 2J(\Delta)$.
To confirm that this choice does not drive the cavity-drain verdict, we repeat the matched-rate comparison of Fig.~2 ($N=6$, cavity drain, $\gamma_{\mathrm{rec}} = \gamma_{\mathrm{deph}} = 1.0$~meV) with a nonzero site diagonal $\varepsilon = \Delta$ added to $\hat{H}$.
Table~\ref{tab:detuning} shows that the rescue efficiency stays at $\eta^*_{\mathrm{rec}} \approx 1$ for every detuning tested, while the dephasing efficiency degrades as $\Delta$ localizes the emitters relative to the cavity, so that the rescue advantage \emph{grows} monotonically with $\Delta$.
The resonant choice used in the main text is therefore the conservative one, and the near-unity rescue transport is independent of the coherent detuning, as expected from the photonic-weight structure of Eq.~\eqref{eq:redfield}.

\begin{table}[h]
\centering
\small
\caption{Robustness of the cavity-drain comparison to a nonzero coherent detuning $\Delta = \varepsilon - \omega_a$ added to $\hat{H}$ ($N=6$, cavity drain, $g=1.5$~meV, $\delta t = 0.5$~meV, $\gamma_{\mathrm{rec}} = \gamma_{\mathrm{deph}} = 1.0$~meV, $\gamma_{\mathrm{lead}} = 0.5$~meV, $t_{\mathrm{mes}} = 30\,\hbar$/meV, 15 disorder realizations). The rescue efficiency is detuning-independent; the advantage over dephasing increases with $\Delta$.}
\label{tab:detuning}
\begin{tabular}{cccc}
\toprule
$\Delta$ (meV) & $\eta^*_{\mathrm{rec}}$ & $\eta^*_{\mathrm{deph}}$ & $\Delta\eta_{\mathrm{peak}}$\\
\midrule
0  & 0.999 & 0.794 & $+20.5\%$\\
5  & 1.000 & 0.611 & $+38.9\%$\\
10 & 1.000 & 0.395 & $+60.5\%$\\
20 & 1.000 & 0.170 & $+83.0\%$\\
\bottomrule
\end{tabular}
\end{table}

\section{Size-scaling of the peak-efficiency gap}
\label{sec:scaling}

\begin{figure}[h]
\centering
\includegraphics[width=0.95\linewidth]{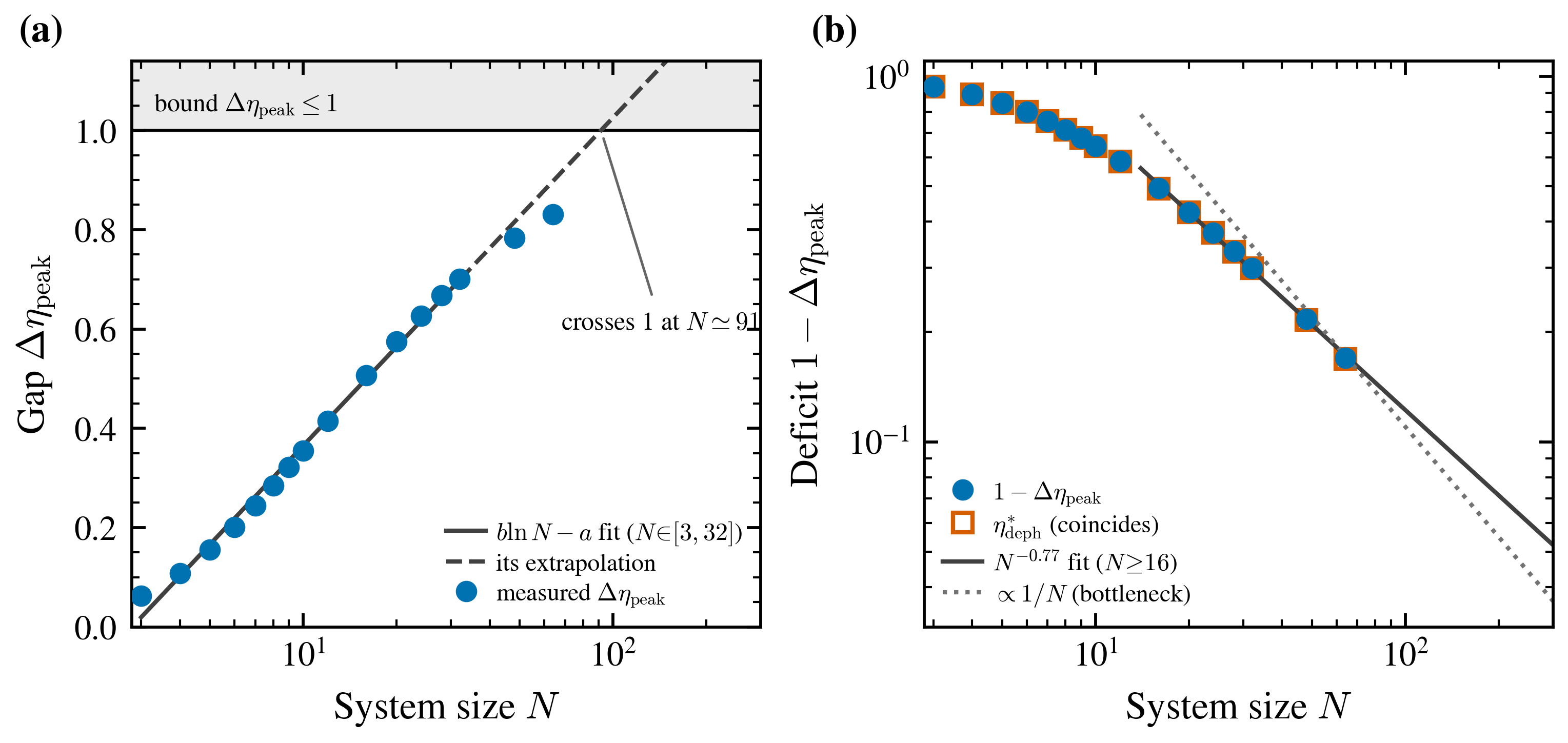}
\caption{Functional form of the peak-efficiency gap (cavity drain, $g=1.5$ meV, $\delta t=0.5$ meV, $T=25\,\hbar$/meV, $N=3$--$64$). (a) Measured $\Delta\eta_{\mathrm{peak}}$ versus $N$ with the logarithmic fit over $N\in[3,32]$ (solid) and its extrapolation (dashed), which crosses the bound $\Delta\eta_{\mathrm{peak}}=1$ (shaded) at $N\simeq91$; the $N=48, 64$ points fall below the extrapolation. (b) The deficit $1-\Delta\eta_{\mathrm{peak}}$ (filled circles; equal to $\eta^*_{\mathrm{deph}}$, open squares) on log-log axes. A power law $N^{-0.77}$ over $N\geq16$ (solid), drifting toward the $\Lambda\sim1/N$ bottleneck slope (dotted). The gap therefore saturates toward complete transport, $\Delta\eta_{\mathrm{peak}}\to1$, with a deficit vanishing as $\sim N^{-1}$.}
\label{fig:scaling_loglaw}
\end{figure}

Subtracting the two optimized efficiency curves of Fig.~1(c) of the main text isolates the rescue advantage $\Delta\eta_{\mathrm{peak}}(N) = \eta^*_{\mathrm{rec}}(N) - \eta^*_{\mathrm{deph}}(N)$. Because $\eta^*_{\mathrm{rec}}\approx1$ throughout, the gap is controlled entirely by the dephasing curve. Over the window $N\in[3,32]$ the gap grows logarithmically, $\Delta\eta_{\mathrm{peak}}(N)\approx b\ln N - a$ with $b = 0.29$, $a = 0.30$ ($R^2 = 0.994$), fitting markedly better than a linear-in-$N$ form ($R^2 = 0.93$) or the dark-fraction form $\propto(N{-}1)/(N{+}1)$ ($R^2 = 0.90$, wrong curvature; both not shown).
The log law is, however, necessarily a finite-window description. Its extrapolation crosses the bound $\Delta\eta_{\mathrm{peak}} = 1$ at $N\simeq91$ [Fig.~\ref{fig:scaling_loglaw}(a)]. The asymptotically consistent statement concerns the \emph{deficit}. Since $\eta^*_{\mathrm{rec}}\approx1$, the gap approaches its ceiling as $\Delta\eta_{\mathrm{peak}} = 1-\eta^*_{\mathrm{deph}}$, and the deficit follows a power law: $1-\Delta\eta_{\mathrm{peak}} = \eta^*_{\mathrm{deph}}\propto N^{-\alpha}$ with $\alpha = 0.77$ ($R^2 = 0.998$ over $N\geq16$) [Fig.~\ref{fig:scaling_loglaw}(b)]. The products $N\eta^*_{\mathrm{deph}} = 9.5,\,10.4,\,10.8$ at $N = 32, 48, 64$ drift toward a constant, i.e.\ the exponent drifts toward the bottleneck value $\alpha = 1$ set by the per-state dark--bright overlap $\Lambda\sim1/N$ --- the same $1/N$ that saturates the no-dissipation baseline ($N\eta_{\mathrm{baseline}}\approx1.08$, constant over the full range). Consistently, the measured $N=48$ and $N=64$ points ($\Delta\eta_{\mathrm{peak}} = 0.783$ and $0.831$) fall below the logarithmic extrapolation ($0.82$ and $0.91$), marking the crossover out of the log window.
Physically, dephasing-mediated dark clearance slows as $\Lambda\sim1/N$ while the rescue rate is pinned by the photonic-weight sum rule and is strictly size-independent, so the advantage approaches complete transport, $\Delta\eta_{\mathrm{peak}}\to1$, with a deficit vanishing as $\sim N^{-1}$ up to slowly varying corrections.
At these sizes the dephasing optimum also moves above the original sweep window: $\gamma^*_{\mathrm{deph}}\approx21.5$ meV at $N=48$ and $64$ (versus $\approx15$ meV at $N=32$), so the $\gamma_{\mathrm{deph}}$ grid was extended from $10$ to $10^{2}$ meV for the $N\ge48$ points (at $N\le32$ the grid-edge effect on $\eta^*_{\mathrm{deph}}$ is below $2\times10^{-3}$, within the disorder error bars).

\section{Eigenstate-resolved evidence for the valve mechanism in the site-$N$ drain}
\label{sec:eigenstate_mechanism}

Figure~2 of the main text shows the manifold-resolved valve dynamics in the cavity-drain configuration.
For completeness, we report the same eigenbasis decomposition in the site-$N$ drain configuration of Table~\ref{tab:regime}, which establishes that the valve is a structural property of the rescue Lindbladian rather than an artifact of any particular drain geometry.

\begin{figure}[h]
\centering
\includegraphics[width=0.95\linewidth]{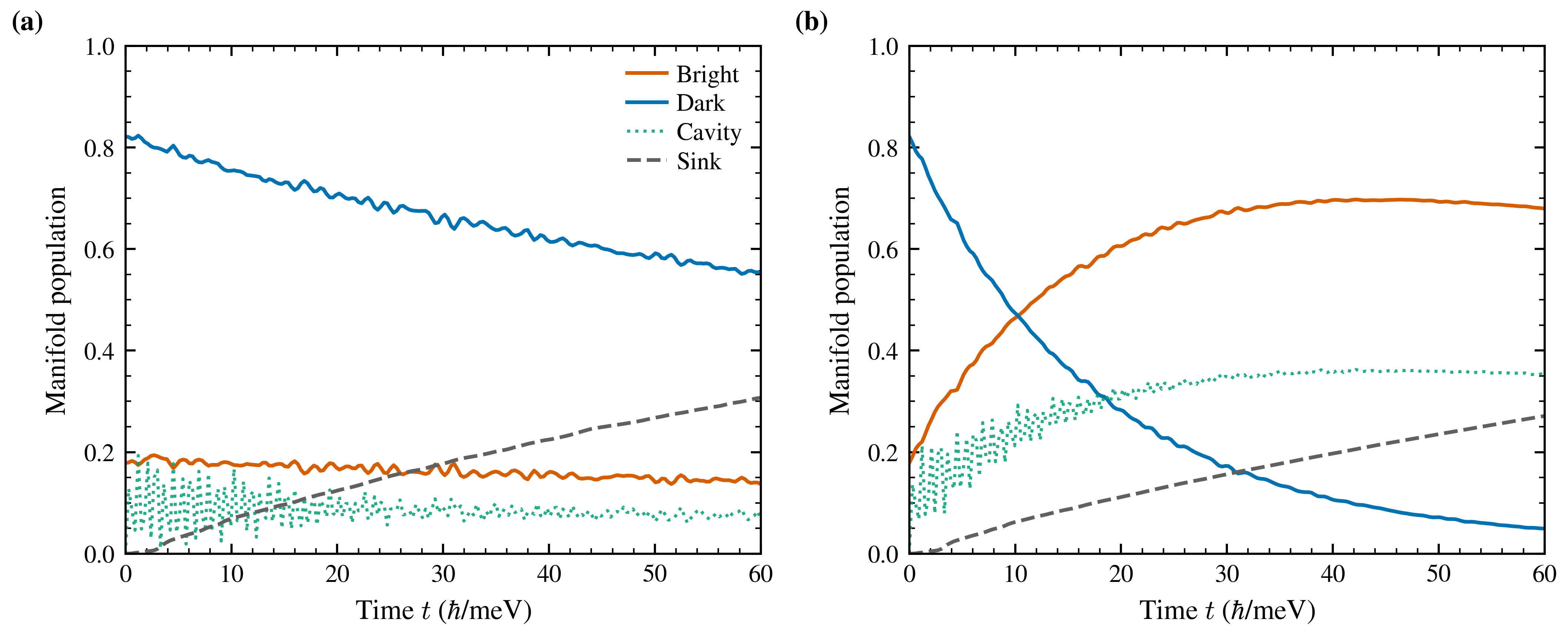}
\caption{Site-$N$ drain analogue of Fig.~2 of the main text. Manifold-resolved population dynamics in the projectors $\hat{P}_{\mathcal{B}}$ (bright), $\hat{P}_{\mathcal{D}}$ (dark), cavity, and sink.
(a) Without rescue ($\gamma_{\mathrm{rec}} = 0$) the dark manifold traps $\approx 80\%$ of the excitation indefinitely.
(b) With rescue ($\gamma_{\mathrm{rec}} = 0.05$ meV) dark population is progressively transferred to a transient bright population and then drained, confirming the valve mechanism.
$N = 6$, $g = 1.5$ meV, $\delta t = 0.5$ meV, site-$N$ drain.}
\label{fig:eigenstates}
\end{figure}

Classifying eigenstates by photonic weight (two largest $w_k$ define $\mathcal{B}$, the remaining $N-1$ define $\mathcal{D}$), Fig.~\ref{fig:eigenstates} traces the manifold projectors $\hat{P}_{\mathcal{B}}$, $\hat{P}_{\mathcal{D}}$ alongside the cavity and sink populations.
Without rescue [Fig.~\ref{fig:eigenstates}(a)], the dark manifold accumulates $\approx 80\%$ of the population and retains it on the simulation window; the bright manifold is depleted and the cumulative drain saturates below $40\%$.
With rescue active [Fig.~\ref{fig:eigenstates}(b), $\gamma_{\mathrm{rec}} = 0.05$ meV], the dark population decays toward zero on the timescale $\gamma_{\mathrm{rec}}^{-1}$ predicted by Eq.~\eqref{eq:closed_form_eta}; the bright manifold maintains a transient, non-monotonic population that funnels into the drain.
The transient bright-manifold population is the universal observable signature of the one-way valve. It appears in both drain geometries and is forbidden for any dark-bright equilibration mechanism.

\section{Detailed-balance derivation of the temperature signature}
\label{sec:T_dependence}

The ratio of absorption to emission rates from Eq.~\eqref{eq:rates} is
\begin{equation}
\frac{\gamma^{(\mathrm{abs})}}{\gamma^{(\mathrm{em})}} = \frac{n(\Delta,T)}{1+n(\Delta,T)} = e^{-\Delta/k_BT},
\end{equation}
which is the Boltzmann factor from detailed balance.
At $T = 0$, the absorption channel is exactly forbidden and the unidirectional valve property holds without correction.
At finite $T$ with $\Delta \gg k_BT$, the reverse process is exponentially suppressed and the main-text predictions apply with corrections of order $e^{-\Delta/k_BT}$.
The crossover regime $k_BT \sim \Delta$ produces the smooth transport-efficiency degradation reported in Fig.~3 of the main text.
For typical molecular polariton detunings $\Delta \sim 10$ meV (cold-mode-resonant), the crossover temperature is $T_{\mathrm{crit}} \sim 100$ K; for intramolecular vibrational detunings $\Delta \sim 100$ meV, the crossover is at $\sim 1000$ K and the rescue is essentially $T$-independent at room temperature.

\begin{figure}[h]
\centering
\includegraphics[width=0.95\linewidth]{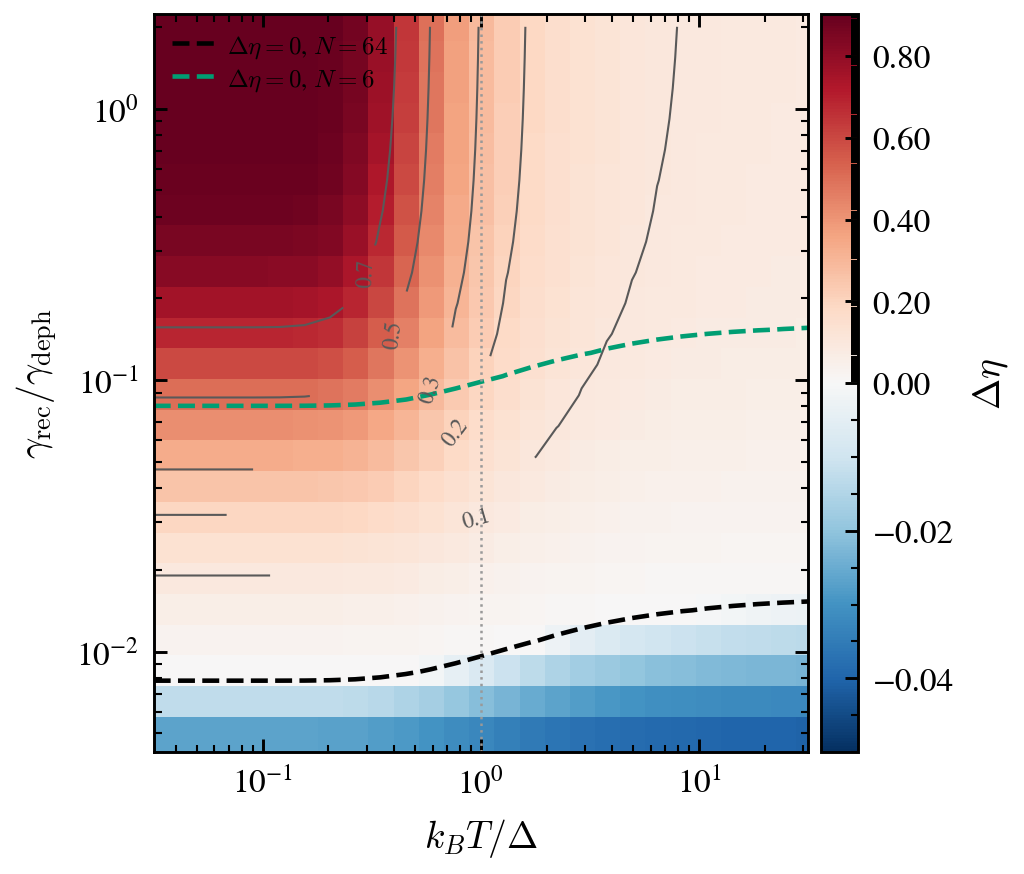}
\caption{$N = 64$ companion to the competition map of main-text Fig.~3(a): efficiency difference $\Delta\eta = \eta_{\mathrm{rescue}} - \eta_{\mathrm{deph}}$ between separate pure-rescue and pure-dephasing propagations vs.\ $k_BT/\Delta$ and $\gamma_{\mathrm{rec}}/\gamma_{\mathrm{deph}}$, at fixed $\gamma_{\mathrm{deph}} = 0.5$~meV (cavity drain, 15 disorder realizations; note the extended ratio axis and the asymmetric color scale).
The $\Delta\eta = 0$ boundary at $N = 64$ (black dashed) lies an order of magnitude below its $N = 6$ counterpart (green dashed, overlaid from the main-text map): as the dark manifold grows, the dephasing benchmark collapses while the sum-rule-protected rescue rate does not, so rescue dominates over nearly the entire rate window at every temperature.}
\label{fig:S_N64map}
\end{figure}

Figure~\ref{fig:S_N64map} repeats the rescue--dephasing competition map of main-text Fig.~3(a) at $N = 64$, with the rate-ratio axis extended down to $\gamma_{\mathrm{rec}}/\gamma_{\mathrm{deph}} = 5\times10^{-3}$.
The $\Delta\eta = 0$ boundary drops by an order of magnitude, from $\gamma_{\mathrm{rec}}/\gamma_{\mathrm{deph}} \approx 0.08$-$0.16$ at $N = 6$ to $\approx 0.008$--$0.015$ at $N = 64$ (both boundaries overlaid in Fig.~\ref{fig:S_N64map}), while the maximum advantage grows from $\Delta\eta \approx 0.34$ to $\approx 0.86$.
This confirms that the fixed-rate ranking inversion of main-text Fig.~3(b,c) holds across the full temperature window. The region where dephasing outperforms rescue shrinks with system size, driven by the collapse of the dephasing benchmark ($\eta_{\mathrm{deph}} = 0.66 \to 0.09$ at $\gamma_{\mathrm{deph}} = 0.5$~meV) rather than by any change in the size-independent rescue rate.
The $T \to 0$ column of the $N = 64$ map agrees with independent GPU-accelerated sweeps of the same protocol to within the disorder standard error ($\lesssim 10^{-3}$ in $\eta$).

\section{QCLE benchmark of the phenomenological Lindblad model}
\label{sec:qcle}

The phenomenological Lindblad framework used throughout the main text was benchmarked against a thermodynamically consistent formulation derived from the quantum-classical Liouville equation (QCLE)~\cite{Dou2015a,Dou2015b,Dou2018}.
This comparison is intended as a consistency check that the phenomenological model reproduces the same steady-state transport and detailed-balance structure obtained from an independent open-system solver; the microscopic bath statistics of the two treatments differ, and the QCLE enters here as a numerical benchmark rather than as a first-principles derivation of the exciton dynamics.
The QCLE friction superoperator is
\begin{equation}
\hat{\hat{\mathcal{L}}}_{\mathrm{QCLE}}\hat{\rho} = -\sum_{m,n}\frac{\Gamma_{mn}}{2\hbar}\Bigl[
\hat{d}^\dagger_m\tilde{\mathbb{D}}_n\hat{\rho} + \hat{d}_m\mathbb{D}^\dagger_n\hat{\rho}
- \hat{d}^\dagger_m\hat{\rho}\,\mathbb{D}_n - \hat{d}_m\hat{\rho}\,\tilde{\mathbb{D}}^\dagger_n + \mathrm{h.c.}\Bigr],
\label{eq:qcle}
\end{equation}
where $\hat{d}_m$ are tunneling operators at lead-coupled sites, $\Gamma_{mn}$ is the system-lead coupling matrix, and the Fermi-weighted bath matrices are constructed in the energy eigenbasis by weighting the matrix element $\langle\varepsilon_j|\hat{d}_n|\varepsilon_k\rangle$ with $f(E_k - E_j, \mu) = [\exp((E_k - E_j - \mu)/k_BT) + 1]^{-1}$ for $\mathbb{D}_n$ and with $1-f$ for $\tilde{\mathbb{D}}_n$.
With this gap convention the superoperator satisfies detailed balance: at equal chemical potentials it relaxes the system to the Gibbs state (verified numerically to four decimal places for $N=4$), and a chemical-potential bias drives the corresponding non-equilibrium steady state.
The transposed convention $f(E_j - E_k, \mu)$ produces an unphysical population-inverted stationary state and must be avoided.

\begin{figure}[h]
\centering
\includegraphics[width=0.85\linewidth]{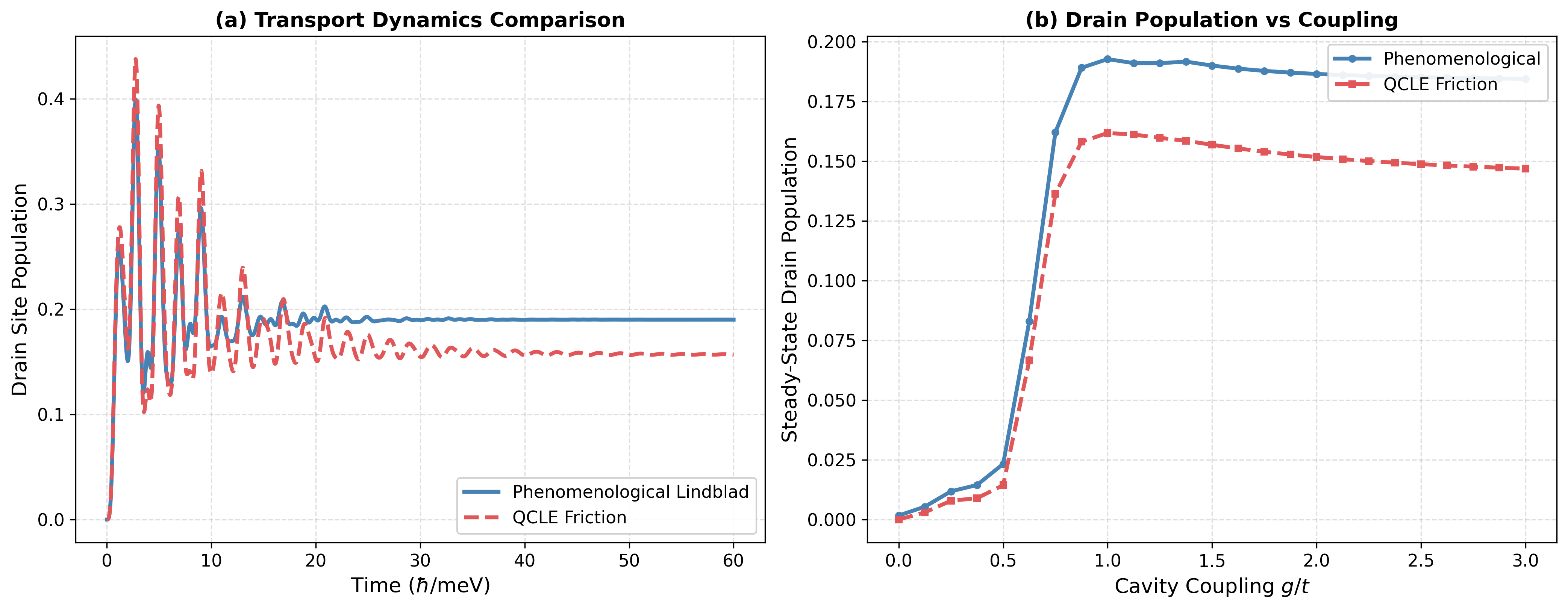}
\caption{Validation of the phenomenological Lindblad model against the QCLE friction formalism for $N=4$. (a) Time evolution of the drain population: both methods track the same coherent transient and relax to steady-state values within $\sim$15\% of each other. (b) Steady-state efficiency as a function of cavity coupling $g$: the onset and the peak (position and height) agree; at larger $g$ the QCLE settles moderately below the phenomenological result.}
\label{fig:qcle}
\end{figure}

Figure~\ref{fig:qcle} compares the QCLE and phenomenological Lindblad time evolution and the steady-state efficiency as a function of cavity coupling $g$ for $N = 4$.
The two methods display the same coherent transient structure, the same efficiency onset, and the same peak position and height (within $2\%$) near $g \approx t$; beyond the peak the QCLE steady state settles moderately ($\lesssim 25\%$ at $g = 3t$) below the phenomenological plateau, as expected when comparing an eigenbasis-constructed (global) Fermi-weighted dissipator with a local phenomenological one.
The phenomenological Lindblad model therefore captures the transport physics relied upon in the main text.

\section{Numerical methods}
\label{sec:numerics_sm}

All numerical results were obtained by direct propagation of the Lindblad equation in the extended Hilbert space (dimension $N+2$) including a decoupled drain sink state.
The Liouvillian was constructed as a $(N+2)^2 \times (N+2)^2$ numpy matrix using column-stacked vectorization,
\begin{equation}
\mathcal{L} = -i(I\otimes \hat{H} - \hat{H}^T\otimes I) + \sum_j\Bigl[\hat{L}^*_j\otimes\hat{L}_j - \tfrac{1}{2}\bigl(I\otimes\hat{L}^\dagger_j\hat{L}_j + (\hat{L}^\dagger_j\hat{L}_j)^T\otimes I\bigr)\Bigr],
\end{equation}
and propagated via direct matrix exponentiation, $\mathrm{vec}(\hat{\rho}(T)) = \exp(\mathcal{L}T)\mathrm{vec}(\hat{\rho}(0))$, using \texttt{scipy.linalg.expm}.
For $N \leq 12$ the Liouvillian dimension does not exceed $14^2 = 196$, and a single propagation takes $\lesssim 10$ ms.
Disorder ensembles use 15--25 independent realizations of $t_i$ sampled from $\mathcal{N}(t,\delta t^2)$, with seeds fixed for reproducibility.

The simulation infrastructure is documented in the project repository.
The cavity-drain regime sweep (Table~\ref{tab:regime} and Fig.~\ref{fig:siteN_phase_diagram}) and the size-scaling sweep (Fig.~1(c) of the main text) each complete in $\sim 60$ s on a single CPU.
The runs for $N>32$ were performed on NVIDIA H100 with 80GB memory using CuPy of CUDA primitives. The $N=32$ GPU run was benchmarked against the corresponding CPU run with $76$x speedup, $1e$-10 tolerance and $<80\%$ GPU utilization. The largest run of $N=96$ for each disorder realization was performed under 50s with $32$GB RAM occupied and 100\% GPU utilization. 

\end{document}